\documentclass[english,lettersize,journal,final]{IEEEtran}
\PassOptionsToPackage{obeyFinal}{todonotes}
\usepackage[T1]{fontenc}
\usepackage{array}
\usepackage{prettyref}
\usepackage{todonotes}
\usepackage{amsmath}
\usepackage{amssymb}
\usepackage{graphicx}
\PassOptionsToPackage{version=3}{mhchem}
\usepackage{mhchem}

\makeatletter

\newcommand{\noun}[1]{\textsc{#1}}
\providecommand{\tabularnewline}{\\}

\usepackage{amsmath,amsfonts}
\usepackage{algorithmic}
\usepackage{algorithm}
\usepackage{array}
\usepackage[caption=false,font=footnotesize,labelfont=sf,textfont=sf]{subfig}
\usepackage{textcomp}
\usepackage{stfloats}
\usepackage{url}
\usepackage{verbatim}
\usepackage{graphicx}
\usepackage{cite}

\usepackage{siunitx}

\usepackage{hyperref}

\DeclareSIUnit\Molar{\textsc{m}}
\DeclareSIUnit\Vol{\micro\meter\cubed}
\DeclareSIUnit\Conc{\textsc{mol.}\per\micro\meter\cubed}
\newrefformat{fig}{Fig.~\ref{#1}}
\newrefformat{tab}{Tab.~\ref{#1}}
\newrefformat{sec}{Sec.~\ref{#1}}
\newrefformat{subsec}{Sec.~\ref{#1}}

\usepackage{tikz}

\newcommand\submittedtext{%
  \footnotesize This work has been submitted to the IEEE for possible publication. Copyright may be transferred without notice, after which this version may no longer be accessible.}

\newcommand\submittednotice{%
\begin{tikzpicture}[remember picture,overlay]
\node[anchor=south,yshift=10pt] at (current page.south) {\fbox{\parbox{\dimexpr0.65\textwidth-\fboxsep-\fboxrule\relax}{\submittedtext}}};
\end{tikzpicture}%
}

\@ifundefined{showcaptionsetup}{}{%
 \PassOptionsToPackage{caption=false}{subfig}}
\usepackage{subfig}
\makeatother

\usepackage{babel}
\begin{document}
\title{Quantitative Aspects, Engineering and Optimization of Bacterial Sensor
Systems}
\author{Florian~Anderl,~\IEEEmembership{Student Member,~IEEE},  Gabriela~Salvadori,~Mladen~Veletic,~Fernanda~Cristina~Petersen and Ilangko~Balasingham,~\IEEEmembership{Senior Member,~IEEE}
\thanks{Florian Anderl is with the Department of Electronic Systems, Norwegian University of Science and Technology, 7491 Trondheim, Norway. }
\thanks{Gabriela Salvadori is with the Intervention Centre, Oslo University Hospital,0424 Oslo, Norway}
\thanks{Mladen Veletic and Ilangko Balasingham are with the Intervention Centre, Oslo University Hospital,0424 Oslo, Norway, and also with the Department of Electronic Systems, Norwegian University of Science and Technology, 7491 Trondheim, Norway.}
\thanks{Fernanda Cristina Petersen is with the Institute of Oral Biology, Faculty of Dentistry, University of Oslo,  0317 Oslo,  Norway}\submittednotice}
\maketitle
\begin{abstract}
Bacterial sensor systems can be used for the detection and measurement
of molecular signal concentrations. The dynamics of the sensor directly
depend on the biological properties of the bacterial sensor cells;
manipulation of these features in the wet lab enables the engineering
and optimization of the bacterial sensor kinetics. This necessitates
the development of biologically meaningful computational models for
bacterial sensors comprising a variety of different molecular mechanisms,
which further facilitates a systematic and quantitative evaluation
of optimization strategies. In this work, we dissect the detection
chain of bacterial sensors from a mathematical perspective from which
we derive, supported by wet-lab data, a complete computational model
for a \emph{Streptococcus mutans}-based bacterial sensor as a case
example. We address the engineering of bacterial sensors by investigating
the impact of altered bacterial cell properties on the sensor response
characteristics, specifically \emph{sensor sensitivity }and \emph{response
signal intensity}. This is achieved through a sensitivity analysis
targeting both the steady-state and transient sensor response characteristics.
Alongside the demonstration of suitability of our methodological approach,
our analysis shows that an increase of sensor sensitivity, through
a targeted manipulation of bacterial physiology, often comes at the
cost of generally diminished sensor response intensity. 
\end{abstract}


\listoftodos{}

\section{Introduction}

\IEEEPARstart{M}{olecular} Communication (MC) is a biology-inspired
communications paradigm that is based on the notion of information
encoded onto molecular carriers\cite{nakanoMolecularCommunication2013}.
It is believed to possess vast potential to revolutionize the health
care field and much initial research is conducted to explore its capabilities\cite{felicettiApplicationsMolecularCommunications2016a}.
MC is the basis for the envisioned \emph{Internet of Bio-Nano Things
}(IoBNT) in which bio-nanomachines are connected in networks and exchange
information over an MC link \cite{akyildizInternetBioNanoThings2015fff}.
IoBNT plays a crucial role in the future of theranostics in form of
implantable bio-nanosensors (BNS) continuously screening health-relevant
parameters and communicating this data in real-time over radio links
to a hub which could be connected to the Internet. A conceivable application
is, for example, the early-detection of bacterial or viral infections
where molecular markers of an ongoing infection are detected. 

Much of research in MC either focuses on the theoretical implications
of MC systems and networks \cite{liuChannelCapacityAnalysis2015,mohamedMolecularCommunicationDiffusion2020,shahmohammadianOptimumReceiverMolecule2012,yilmaz3DChannelCharacteristics2014a}
or quantitative information extraction from biological systems \cite{veleticMolecularCommunicationModel2019,barrosMolecularCommunicationsViral2021}.
As MC ultimately aims at engineering and prototyping practical solutions,
selected works study feasible MC components. The survey \cite{kuscuTransmitterReceiverArchitectures2019},
for example, explores the requirements, communication-theoretic properties
and possible architectures for MC transmitter (Tx) and receiver (Rx)
components. It distinguishes between two main classes of conceivable
architectures: nanomaterial-based and biology-based designs. An example
for the former category is the graphene-based bioFET MC receiver design
\cite{kuscuFabricationMicrofluidicAnalysis2021}. In contrast, the
latter category leverages synthetic biological elements, e.g. , human
or bacteria cells that can be genetically engineered for a specific
purpose. Here, bacteria are especially relevant due to the relative
ease of genetically engineering them for specific purposes which make
bacteria invaluable for various scientific, health-care related and
industrial applications. 

Consequently, MC has acknowledged this potential in several works.
Examples include a concept where engineered bacteria are repurposed
as transceivers with different operational modes in an MC link \cite{unluturkGeneticallyEngineeredBacteriaBased2015a}.
Moreover, an MC architecture with proton pumping bacteria was recently
investigated \cite{grebensteinMolecularCommunicationTestbed2019},
and a bacterial receiver prototype using the rhamnose operon in a
microfluidic environment was presented alongside a computational model\cite{amerizadehBacterialReceiverPrototype2021}. 

MC borrows the utility of bacteria as receivers from a key area of
application for synthetic bacteria: the \emph{sensing }of analytes.
The deployment of bacteria as biosensors for a variety of substances
is a well-researched field in microbiology and bio-engineering. Many
works explore the principles and techniques for engineering bacterial
biosensors \cite{vandermeerBacterialSensorsSynthetic2010,vandermeerWhereMicrobiologyMeets2010,jungBacterialTransmembraneSignalling2018},
and investigate various areas of applications for bacterial biosensors,
from environmental monitoring \cite{teconBacterialBiosensorsMeasuring2008,wuWholeCellBiosensorPointofCare2021,woutersenAreLuminescentBacteria2011}
to medical applications \cite{strussPaperStripWhole2010,millerDetectionQuorumSensingMolecules2020,riglarEngineeringBacteriaDiagnostic2018,tannicheEngineeredLiveBacteria2023,changMicrobiallyDerivedBiosensors2017}.

While much of the related literature describes a more qualitative
approach, only few works attempt to derive quantitive measures for
bacterial biosensors \cite{vandermeerIlluminatingDetectionChain2004b}.
Numerous computational models for dynamics arising from specific bacterial
mechanisms have been published \cite{emereniniMathematicalModelQuorum2015,dockeryMathematicalModelQuorum2001,meyerDynamicsAHLMediated2012,anguigeMathematicalModellingTherapies2004a,gustafssonCharacterizingDynamicsQuorumSensing2005,haustenneModelingComRSSignaling2015}.
Nevertheless, to the knowledge of the authors none addresses the general
dynamics of bacterial systems with focus on its sensor properties.
A lack of a computational framework that describes the sensing dynamics
of bacterial sensors prevents us from proceeding with a systematic
investigation of how the dynamics can be altered to adhere to specific
requirements of the engineered bacterial sensor. 

Most conceivable bacterial biosensors rely on the same key processes
and biochemical principles and thus have similar structures. This
similarity is useful for a computational analysis of these key processes
and the resulting qualitative and quantitative dynamics of bacterial
sensors in general. This analysis further allows for a systematic
investigation of how these dynamics can be altered to adhere to specific
requirements on the bacterial sensor. In this paper, we therefore
conduct a computational analysis of bacterial sensors and how bacteria
can be engineered and optimized catering to the requirements of building
actual bacteria-based components for MC. We examine the general components,
integral features and critical processes of bacterial sensors with
an emphasis on sensing and detection. Through this, we summarize the
mechanisms that underlie general bacterial sensors and derive corresponding
mathematical models. Using methods from biochemical systems theory
(BST) we analyze the steady-state and transient dynamics of a \emph{generic}
bacterial sensor. We then proceed to derive expressions for the sensitivity
and signal intensity of the sensor. For these measures we then conduct
a sensitivity analysis and thereby identify critical parameters of
the model and quantify how changes in these affect the model dynamics.
For example, similar to an antenna in a telecommunications link or
a radar dish, a bacteria-based MC sensor- or receiver should exhibit
a measure for gain. 

As a case study, we present a concrete example of utilizing the proposed
general computational framework for building a computational model
for analyzing a bacterial sensor based on experimentation with the
bacterial species \emph{Streptococcus mutans}. As basis for a reporter,
we use the \emph{luc} gene encoding the enzyme luciferase from the
firefly \emph{Photinus pyralisfirefly} (further referrer to as\emph{
lucFF}), which is inserted into the genome of a bacterial sensor strain.
These modified bacteria produce the luciferase enzyme, which upon
addition of luciferin results in the emission of a light signal proportional
to a chemical target signal, functioning as sensors. This allows,
for instance, integration with a photodetector which further facilitates
integration with electronic elements. 

The remainder of the paper is structured as follows. \prettyref{sec:Bacteria-and-Biosensors}
discusses important aspects of bacteria as biosensors and presents
necessary biological background information. \prettyref{sec:Theoretical-Framework}
provides the theoretical framework which we employ for our analysis.
\prettyref{sec:Dissecting-the-Bacterial} analyzes the discrete processes
in a bacterial sensor and summarizes mathematical modeling approaches
for the mechanisms. In \prettyref{sec:Bacterial-Sensor-Models}, we
then present a mathematical model for a \emph{Streptococcus mutans-}based
sensing system and compare this model to our empirical wet-lab data.
\prettyref{sec:Optimization-of-Bacterial-Sensors} presents a systematic
scheme for analyzing a bacterial sensor's performance. Based on this,
we comment on strategies for optimization of bacterial sensors. Finally,
\prettyref{sec:Conclusion} concludes the paper and outlines potential
research building on this work.

\section{Bacteria as Biosensors \label{sec:Bacteria-and-Biosensors} }

\begin{figure}[t]
\centering{}\includegraphics[width=0.95\columnwidth,height=0.3\textheight,keepaspectratio]{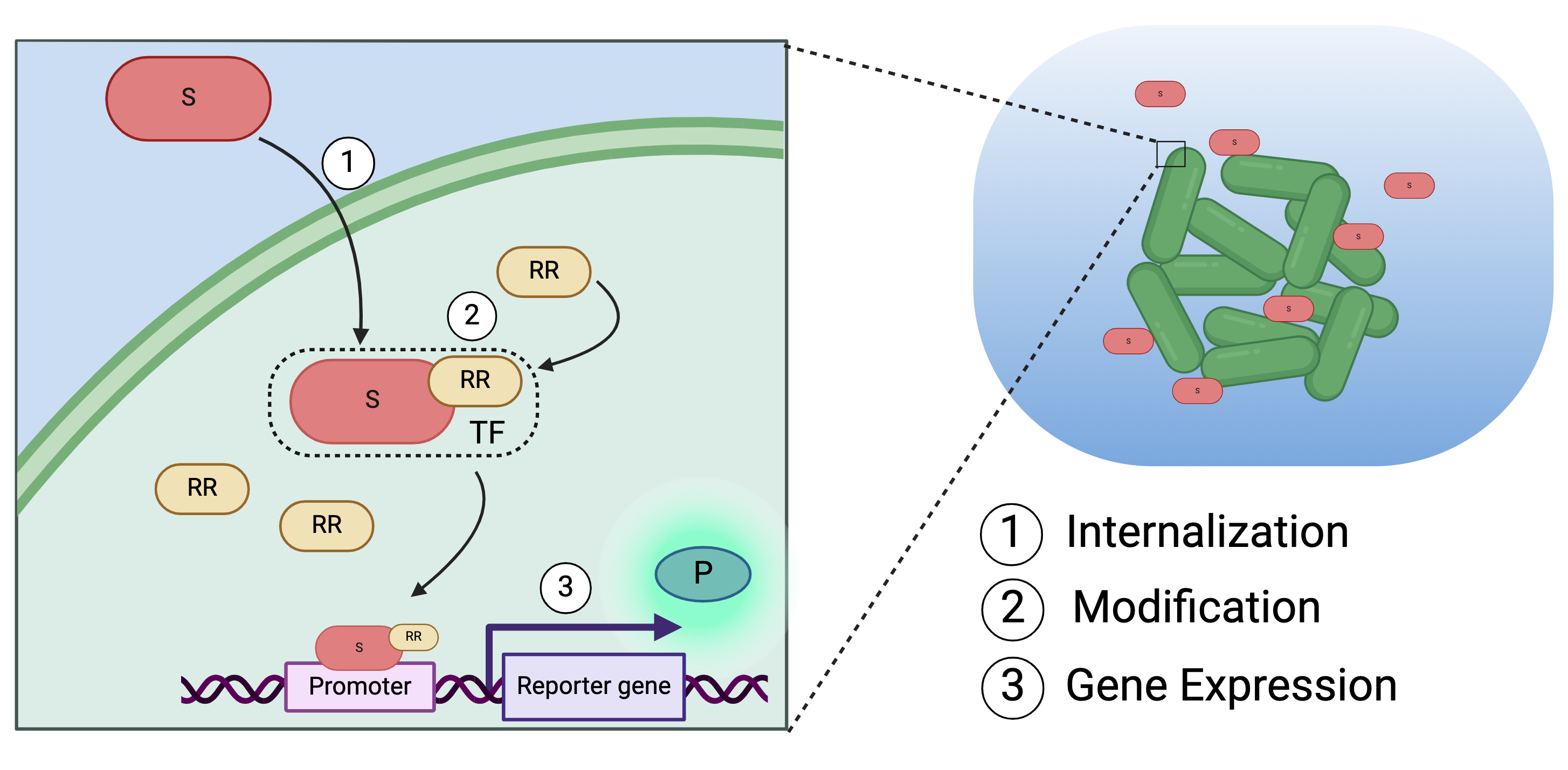}\caption{\label{fig:Bacterial-sensor-mechanism}{\scriptsize{} }Overview of
general bacterial sensor mechanism. Molecular signal $\mathsf{S}$,
response regulator $\mathsf{RR}$, transcription factor $\mathsf{TF}$
and reporter $\mathsf{P}$. (Illustration was created with \emph{BioRender.com}). }
\end{figure}

The general structure of a bacterial sensor is depicted in \prettyref{fig:Bacterial-sensor-mechanism}.
The bacteria's molecular response pathway is activated by a target
analyte or effector eliciting the expression of reporter genes which,
in the genome, are inserted downstream of the regulated primary gene.
The reporter proteins resulting from this gene expression or rather
their specific activity then are a proxy for the analyte concentration.
This mechanism is governed by several distinct sub-processes emphasized
in \prettyref{fig:Bacterial-sensor-mechanism}: \emph{internalization,
modification }and \emph{gene expression. }

Internalization describes the uptake process of the extracellular
signal into the sensing cell either directly or by effect as is the
case in, for example, phosphorylation. In this case, the external
signal binds to receptors on the cell surface which elicits a response
inside the cell. Subsequent to uptake, the internalized analyte may
undergo several modification steps involving a \emph{response regulator
}(RR). Finally, the resulting \emph{transcription factor} (TF) then
regulates gene expression. These processes are detailed further below
in \prettyref{sec:Dissecting-the-Bacterial}.

Various reporter mechanisms exist, most prominently: colorimetry,
fluorescence (FL) and bioluminescence (BL). Colorimetry is based on
assay color changing reactions elicited by, for example, beta-galactosidase
resulting from the expression of the \emph{lacZ }reporter gene. FL
and BL rely on chemical reactions producing light at specific wavelengths.
FL resulting from expression of genes like \emph{gfp, egfp} or \emph{mCherry
}requires excitation from an external light source. In contrast, BL
functions without external activation which makes it especially relevant
for applications in which the bacterial sensor component is integrated
into an electronic device. The most important BL reporter proteins
are the bacterial luciferase variants LuxAB, LuxCDABE and firefly
luciferase LucFF. Various aspects influence the choice of reporter
protein: substrate availability (for example, BL relies on oxygen
and luciferin), sensitivity and, crucially, harvestability. A comprehensive
overview of reporters is given in \cite{vandermeerBacterialSensorsSynthetic2010}.

We further distinguish between single-cell and population-based bacterial
sensors. The former exhibits a cellular granularity since the response
of single cells is considered \cite{ramalhoSingleCellAnalysis2016,teconInformationSinglecellBacterial2006b}.
In contrast, in the population-based approach the response of the
entire population of the sensing strain is measured. The stochasticity
inherent in molecular mechanisms (especially relevant at low analyte
concentrations) and gene expression results in generally non-deterministic
single-cell bacterial sensor dynamics. This implies heterogenous reporter
expression levels across a sensor population. This heterogeneity can
be measured through microscopy imaging and be advantageous for certain
sensing applications as it can potentially resolve spatial information. 

In this paper, we restrict our treatment to an analysis of whole-cell
bacterial sensors on population-basis deployed in an aqueous assay.
This implies a homogeneous concentration of a target analyte in the
assay and each viable bacterial cell being exposed to the same concentration
of target analyte. We assume further that each viable bacterial cell
contributes equally to the overall sensor response, i.e., we ignore
the natural heterogeneity of gene expression across a bacterial population.
We justify this by our aiming at deriving a model for the \emph{mean
bacterial cell response},\emph{ }as discussed in \prettyref{subsec:Streptococcus-mutans-as}. 

We further distinguish between two general sensing scenarios: 

\paragraph{Scenario 1 - Abundance of Analyte}

In this case, the analyte concentration in the sensing cavity is sufficiently
high so that it is not influenced or noticeably diminished by the
sensor dynamics itself. This could be the case if there exists a constant
supply of analyte into the sensing cavity which is a reasonable assumption
for cases of e.g., acute infection with a virulent bacterial strain
that is in the stationary phase of its growth cycle. If the virulent
strain emits a specific detectable signal (e.g., through \emph{Quorum
Sensing}), we can assume an abundant signal supply. In this case,
the target analyte enters the computational model describing the sensor
as \emph{constant} \emph{independent variable}. As a result, a bacterial
sensor cell reaches a non-zero steady-state response level.

\paragraph{Scenario 2 - Scarcity of Analyte}

When the effector signal is at a low concentration and depleted without
experiencing a positive influx it must be included in the computational
model as an additional equation as the bacterial sensor \emph{itself}
affects the concentration of analyte. 

In \prettyref{sec:Optimization-of-Bacterial-Sensors}, we use \emph{Scenario
1} and\emph{ Scenario 2} for a systematic steady-state and dynamic
optimization analysis, respectively. 

\section{Theoretical Framework \label{sec:Theoretical-Framework}}

Bacterial sensors can be analyzed as biochemical systems at different
levels of abstraction. Low concentrations of involved biochemical
species mandate stochastic and particle-based models accounting for
the probabilistic nature of molecular interactions. Monte-Carlo simulations
allow for the synthesis of deterministic trajectories from the stochastic
models. Alternatively, direct deterministic models based on reaction-rate
equations (RREs) are often a simpler approach while retaining the
capability of capturing the system dynamics accurately. 

In this section, we give a brief overview of BST as it provides methods
that are crucial for our analysis of a bacterial sensor. In the following
section, we then outline dynamics of various bacterial sensors and
subsequently apply the BST methodology to the derived models of bacterial
sensors to provide a systematic analysis of their performance, thus
revealing potential starting points for their optimization.

\subsection{Biochemical Systems Theory \label{subsec:Biochemical-Systems-Theory} }

BST studies methods for mathematical modeling of biochemical systems
\cite{voitBiochemicalSystemsTheory2013}. It does so by deriving
simplistic approximative models that allow for facilitated mathematical
analysis while retaining the dynamics observed in the data.\emph{
}The analysis of a biochemical system generally begins with the identification
of structural relationships in the studied system which can be comprehensively
illustrated by graphical representations (as done in \prettyref{sec:Dissecting-the-Bacterial},
cf. \prettyref{fig:arrow_diag_bac_sens}). The dynamics of a system
with various constituent species $X_{1},X_{2},...,X_{n}$ are generally
described by a set of ordinary-differential-equations (ODEs) parametrized
by $\mathbf{\Theta}_{R}=\{\theta_{1},...,\theta_{F}\}$ as
\begin{align}
\dot{X}_{1} & =v_{1}^{+}(X_{1},...,X_{n},...,X_{n+m};\mathbf{\Theta}_{R})\nonumber \\
\vdots & \hspace{1cm}-v_{1}^{-}(X_{1},...,X_{n},...,X_{n+m};\mathbf{\Theta}_{R})\nonumber \\
\dot{X}_{n} & =v_{n}^{+}(X_{1},...,X_{n},...,X_{n+m};\mathbf{\Theta}_{R})\nonumber \\
 & \hspace{1cm}-v_{n}^{-}(X_{1},...,X_{n},...,X_{n+m};\mathbf{\Theta}_{R})\label{eq:BST_ode_general}
\end{align}
where the flux terms $v_{i}^{+/-},i=1,...,n$ are either positive
fluxes adding to a pool of species $X_{i}$ or negative when depleting
it. These fluxes are in general functions of the system's $n$ internal
state variables $\mathbf{X}_{\mathsf{int}}=\left[X_{1},...,X_{n}\right]^{T}$
and $m$ external variables $\mathbf{X}_{\mathsf{ext}}=\left[X_{n+1},...,X_{n+m}\right]^{T}$.
The fluxes specifically depend on a subset of variables identified
in the structural analysis of the biochemical system under investigation.
The flux terms can assume various different functional forms: from
basic RREs to terms describing complex mechanisms as, for example,
\emph{Michaelis-Menten-kinetics} or \emph{Competitive Inhibition}. 

It is mathematically convenient to employ an universal form for representing
the fluxes $v_{i}^{+/-}$ regardless of the dynamics governing the
specific case and this form is provided by the\emph{ Power-Law-Formalism.
}As in a classical mass-action derived RRE, a single flux is modeled
as a product of a kinetic rate constant $\gamma$ and involved species
among which each are raised to a power of a specific kinetic rate
(or kinetic orders) $f$, i.e., 
\begin{equation}
v_{i}^{+/-}=\gamma_{i}\prod_{j=0}^{n+m}X_{j}^{f_{ij}},\quad i=1,...,n.\label{eq:power_law_term}
\end{equation}

In \eqref{eq:BST_ode_general}, a flux $v_{i}^{+/-}$ can consist
of several distinct processes (chemical reactions, molecular interactions,
etc.), i.e., $v_{i}^{+/-}=\sum_{k=1}^{T_{i}}v_{ki}^{+/-}$. The choice
of representing a biochemical system with aggregated flux terms $v_{i}^{+/-}$or,
alternatively, distinguishing single process rates $v_{ki}^{+/-},k=1,...,T_{i}$
yields either an \emph{S-System }or a \emph{General-Mass-Action }(GMA)
system, respectively.\footnote{For the sake of completeness, a GMA system is given by
\[
\dot{X}_{i}=\sum_{k=1}^{T_{i}}\pm\gamma_{ik}\prod_{j=1}^{n+m}X_{j}f^{ikj},\quad i=1,...,n
\]
All positive (negative) terms in the above equation are aggregated
into the first (second) term on the right hand side of \eqref{eq:canonical_s_system_form}.
As the name implies, GMA systems are generalized mass action\emph{
}formulations and allow for non-integer valued kinetic orders. In
that sense, allowing the kinetic orders to assume any real number
renders a GMA model more flexible. }Consequently, the canonical BST formulation of an S-system with $n$
dependent and $m$ independent variables has the format \cite{savageauIntroductionSsystemsUnderlying1988}
\begin{equation}
\dot{X}_{i}=\alpha_{i}\prod_{j=1}^{n+m}X_{j}^{g^{ij}}-\beta_{i}\prod_{j=1}^{n+m}X_{j}^{h^{ij}},\quad i=1,...,n.\label{eq:canonical_s_system_form}
\end{equation}

Rate terms of the form \eqref{eq:power_law_term} are in general \emph{approximations}
to the real dynamics of the specific process. Consequently, the parameters
$\mathbf{\Theta}_{S}=\left\{ g_{ij},h_{ij},\alpha_{i},\beta_{i}\right\} ,i=1,...,n;j=1,...n+m$
of an S-system can be derived from the base system as \cite{savageauIntroductionSsystemsUnderlying1988}
\begin{align}
g_{ij},h_{ij} & =\left.\frac{\partial\log\left(\sum_{k=1}^{T_{i}}v_{k}^{+,-}\right)}{\partial\log X_{j}}\right|_{\mathbf{X=X}_{0}}\nonumber \\
 & =\left.\frac{\partial\left(\sum_{k=1}^{T_{i}}v_{k}^{+,-}\right)}{\partial X_{j}}\right|_{\mathbf{X=X}_{0}}\frac{X_{j}}{\left(\sum_{k=1}^{T_{i}}v_{k}^{+,-}\right)},\label{eq:S_system_params_kinetic_rates}\\
\alpha_{i},\beta_{i} & =v_{i}^{+,-}\left(\mathbf{X}_{0}\right)\prod_{j=1}^{n+m}X_{j}^{-g_{ij},h_{ij}}.\label{eq:S_system_params_rate_consts}
\end{align}
In \eqref{eq:S_system_params_kinetic_rates} and \eqref{eq:S_system_params_rate_consts}
$\mathbf{X}_{0}=\left[X_{1_{0}},...,X_{n_{0}}\right]^{T}$ is the
\emph{operating point }at which the S-system is constructed. At this
operating point, equality holds between the base biochemical system
and its corresponding S-system form.

We introduce S-systems here since, in \prettyref{sec:Optimization-of-Bacterial-Sensors},
we make use of their convenient mathematical properties which facilitate
analytic steady-state analysis. We briefly outline these mathematical
properties in the following.

\subsection{Steady-State Analysis \label{subsec:Steady-State-Analysis}}

The canonical S-system form has convenient mathematical properties
which allow for the direct analytic calculation of its steady-state
solution in the log-space as \cite{savageauIntroductionSsystemsUnderlying1988}
\begin{equation}
\mathbf{y}_{\mathrm{int}}=\mathbf{L}\mathbf{y}_{\mathrm{ext}}+\underset{\mathbf{m}}{\underbrace{\mathbf{M}\mathbf{b}}}\label{eq:S_system_steady_state_eq}
\end{equation}
where the matrix $\mathbf{L}\in\mathbb{R}^{n\times m}$ contains the
\emph{log-gains} which relate changes in logarithms of independent
(external) variables in $\mathbf{y}_{\mathsf{ext}}\in\mathbb{R}^{m\times1}$
to changes in logarithm of dependent (internal) variables $\mathbf{y}_{\mathsf{int}}\in\mathbb{R}^{n\times1}$. 

Entries of $\mathbf{L}$, $\mathbf{M}$ and $\mathbf{b}$ are composed
of functions of the S-system parameters $\mathbf{\Theta}_{S}$ and
are thus functions of $\mathbf{\Theta}_{R}$, i.e., the parameters
of the original RRE following \eqref{eq:S_system_params_kinetic_rates}
and \eqref{eq:S_system_params_rate_consts}. Eq. \eqref{eq:S_system_steady_state_eq}
implies that the independent variables in $\mathbf{y}_{\mathrm{int}}$
are \emph{linear }functions of the independent variables in $\mathbf{y}_{\mathrm{ext}}$
with the slope of the resulting line given by the entries $l_{ij}=\frac{\partial X_{i}}{\partial X_{j}}\frac{X_{j}}{X_{i}},i=1..,n;j=1,...,m$
of $\mathbf{L}$ and the constant offset given by element of the vector
$\mathbf{m}$ resulting from the matrix-vector product $\mathbf{M}\mathbf{b}$.

We use the form of \eqref{eq:S_system_steady_state_eq} in the application
to a bacterial sensor model in \prettyref{sec:Optimization-of-Bacterial-Sensors}
to directly obtain steady-state figures that can be linked to general
sensors properties. 

Considering the optimization and engineering of bacterial sensors
as the main goal of this work, we then proceed with an analysis of
how these sensor properties depend on the sensor model parameters.
Formally, we achieve that through a \emph{sensitivity analysis.}

\subsection{Steady-State Sensitivities \label{subsec:Steady-State-Sensitivities}}

From \eqref{eq:S_system_steady_state_eq} one can directly obtain
the sensitivities of the steady-state solutions $\mathbf{y}_{i}$
with respect to the S-system parameters $\mathbf{\Theta}_{S}$ and
in extension of $\mathbf{\Theta}_{R}$. The \emph{relative} sensitivities
$S_{l_{ij},\theta_{k}},S_{m_{i},\theta_{k}}$ are calculated for $l_{ij}\in\mathbf{L}$
and $m_{i}\in\mathbf{m}$ using the chain-rule as \cite{savageauConceptsRelatingBehavior1971}
\begin{align}
S_{l_{ij},\theta_{k}} & =\frac{\partial l_{ij}}{\partial\theta_{k}}\left|\frac{\theta_{k}}{l_{ij}}\right|=\sum_{f\in\tilde{\Theta}_{S}}\frac{\partial l_{ij}}{\partial f}\frac{\partial f}{\partial\theta_{k}}\times\left|\frac{\theta_{k}}{l_{ij}}\right|,\label{eq:log_gain_sensitivities}\\
S_{m_{i},\theta_{k}} & =\frac{\partial m_{j}}{\partial\theta_{k}}\left|\frac{\theta_{k}}{m_{j}}\right|=\sum_{f\in\Theta_{S}}\frac{\partial m_{i}}{\partial f}\frac{\partial f}{\partial\theta_{k}}\times\left|\frac{\theta_{k}}{m_{i}}\right|,\label{eq:m_sensitivities}\\
 & \quad\theta_{k}\in\mathbf{\Theta}_{R},\tilde{\mathbf{\Theta}}_{S}\subseteq\mathbf{\Theta}_{S},i=1,...n,j=1,...,m.\nonumber 
\end{align}
Eqs. \eqref{eq:log_gain_sensitivities} and \eqref{eq:m_sensitivities}
evaluate the \emph{percentage} changes in $\mathbf{L}$ and $\mathbf{m}$
due to infinitesimal percentage changes in $\theta_{k}\in$ $\mathbf{\Theta}_{R}$.
We use these measures to investigate the contribution of biological
parameters to the steady-state behavior of bacterial sensors.

\subsection{Dynamics Sensitivities \label{subsec:Dynamics-Sensitivities}}

In addition to the steady-state treatment, we conduct a \emph{dynamic
sensitivity analysis }for bacterial sensor models. Here we assume
that the investigated biochemical system has \emph{no} independent
variables as the kinetics of the analyte state variable are now affected
by the system dynamics.

With $\mathbf{f}_{c}\in\mathbb{R}^{n\times1}$ comprising elements
$\partial\dot{X_{i}}/\partial\theta_{k},i=1,...,n,\theta_{k}\in\mathbf{\Theta}_{R}$,
$\mathbf{J}\in\mathbb{R}^{n\times n}$ as the Jacobian with entries
$\partial\dot{X}_{i}/\partial X_{j},i,j=1,...,n$ and dynamic sensitivity
vector $\mathbf{z}_{k}\in\mathbb{R}^{n\times1}$ with entries $\partial X_{i}/\partial\theta_{k}$\cite{dickinsonSensitivityAnalysisOrdinary1976},
\begin{equation}
\dot{\mathbf{z}}_{k}=\mathbf{f}_{c}+\mathbf{J}\mathbf{z}_{k},\quad k=1,...,F\label{eq:dynamic_sensitivities_diff_eq}
\end{equation}
denote the differential equations of the dynamic sensitivities with
respect to parameter $\theta_{k}$. An ODE system consisting of $(F\cdot n)$
equations of the form \eqref{eq:dynamic_sensitivities_diff_eq} can
then be solved numerically to yield the final dynamic sensitivities
$\mathbf{z}$ that describe the temporal evolution of the sensitivities.
This provides insight into how different phases of the time-course
dynamics of a variable $X_{i}$ responds to changes in parameters
$\mathbf{\Theta}_{R}$. 

One can further include the initial conditions of $\mathbf{X}$ into
$\mathbf{\Theta}_{R}$ to investigate their effects on the system
dynamics. In fact, by calculating the \emph{relative dynamic sensitivity}
with respect to the initial condition $X_{j}(0)$ of the \emph{dependent
}variable\emph{ }$X_{j}$, i.e., 
\begin{equation}
S_{i,X_{j}(0)}=\frac{\partial X_{i}}{\partial X_{j}(0)}\frac{X_{j}(0)}{X_{i}},\quad i,j=1,...,n\label{eq:rel_dynamic_sensitivity_init_cond}
\end{equation}
a measure is obtained which is very similar to the \emph{dynamic log-gain}
\cite{shiraishiEfficientMethodCalculation2005}.\emph{ }The dynamic
log-gain can be used to evaluate the influence of changes in independent
variables on the time-course of the investigated state-variable. Eq.
\eqref{eq:rel_dynamic_sensitivity_init_cond}, in contrast, assumes
$X_{j}$ to be dependent, hence the sensitivity with respect to its
initial condition is measured. 

We can then proceed to calculate the sensitivities of $S_{i,X_{j}(0)}$
with respect to other model parameters $\theta_{k}\in\mathbf{\Theta}_{R}$
by solving the following differential equation (for derivation see
Appendix \ref{sec:Derivation-of-Dynamic}):
\begin{align}
\frac{\mathrm{d}}{\mathrm{d}t}\left(\frac{\partial S_{i,X_{j}(0)}}{\partial\theta_{k}}\right) & =-\left(\frac{\partial S_{i,X_{j}(0)}}{\partial\theta_{k}}X_{i}+S_{i,X_{j}(0)}\frac{\partial X_{i}}{\partial\theta_{k}}\right)\frac{\dot{X_{i}}}{X_{i}^{2}}\nonumber \\
 & \hspace{-0.5cm}+X_{j}(0)\frac{\partial X_{i}}{\partial X_{j}(0)}\left(2X_{i}^{-3}\dot{X_{i}}-X_{i}^{-2}\frac{\partial\dot{X_{i}}}{\partial\theta_{k}}\right).\label{eq:deq_dyn_log_gain_sensitivity}
\end{align}
Eq. \eqref{eq:deq_dyn_log_gain_sensitivity} constitutes the dynamic
sensitivity analysis equivalent of \eqref{eq:log_gain_sensitivities}.
As it shares terms with \eqref{eq:dynamic_sensitivities_diff_eq},
it can be integrated with $\dot{\mathbf{z}}_{k}$ into an extended
ODE system.

\section{Dissecting the Bacterial Sensor Detection Chain \label{sec:Dissecting-the-Bacterial}}

\begin{figure}
\centering{}\includegraphics[width=0.95\columnwidth,height=0.15\textheight,keepaspectratio]{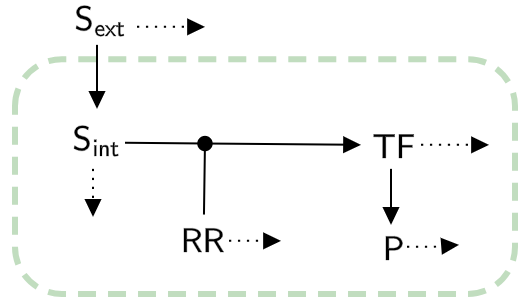}\caption{\label{fig:arrow_diag_bac_sens}Structural diagram of a bacterial
sensor. The dotted arrows indicate that all species are subject to
degradation. }
\end{figure}
The signal transduction pathway depicted in \prettyref{fig:arrow_diag_bac_sens}
shall serve as reference for the analysis of the various distinct
steps in the detection chain of a bacterial sensor and the respective
computational models. In the most general sense, the target analyte
$\textsf{S}_{\textsf{ext}}$ is \emph{internalized} by a sensor cell
and then potentially undergoes \emph{modification} in interaction
with response regulator $\textsf{RR}$. The resulting transcription
factor $\textsf{TF}$ subsequently drives the \emph{expression} of
the gene coding for the reporter protein $\textsf{P}$. Importantly,
the internalization of the effector $\textsf{S}_{\textsf{ext}}$ can
either occur directly or indirectly by effect only. This is the case
of the \emph{phosphorylation} of an internal response regulator caused
by an effector-bound cell surface receptor. We dissect the detection
chain into its distinct processes and analyze their mathematical models.
These models are initially provided as ordinary chemical reaction
rate equations, since models of this form are most intuitive and generally
a good starting point for a computational analysis. We note that parts
of this section do not present new findings but collects information
from various sources. We deem it necessary to proceed in this fashion
as the presented information is crucial to comprehensively discussing
bacterial sensor dynamics. 

\subsection{Internalization }

The first step in the bacterial sensor detection chain is the uptake
of the target analyte $\mathsf{S_{ext}}$ into the bacterial sensor
cells (\prettyref{fig:arrow_diag_bac_sens}). The variety of bacteria
strains exhibit a multitude of different uptake mechanisms that vary
in their biochemical dynamics. The transport mechanisms can be categorized
into the following groups:

\subsubsection*{Diffusion}

In many bacteria nutrients or molecular signals \emph{diffuse} into
the bacterial cells across the cell wall. This is a passive process
and requires no energy expenditure by the cell. This process can generally
be described by Fick's law of diffusion across a membrane as \cite{enderleChapterCompartmentalModeling2012,meyerDynamicsAHLMediated2012}
\begin{align}
\frac{\mathrm{d}\mathsf{S}_{\mathsf{int}}}{\mathrm{d}t} & =-q_{D}\left(\frac{\mathsf{S}_{\mathsf{int}}}{V_{\mathrm{i}}}-\frac{\mathsf{S}_{\mathsf{ext}}}{V_{\mathrm{e}}}\right)-\delta_{\mathsf{s}}\mathsf{S_{int}}\label{eq:Fick_int_rre}
\end{align}
where $\mathsf{S_{ext}}$ and $\mathsf{S_{int}}$ denote the external
and internal amount of target analyte, respectively, $V_{\mathrm{i}}$
and $V_{\mathrm{e}}$ are the intracellular and extracellular volumes
and $\delta_{\mathsf{s}}$ is the degradation rate constant of $\mathsf{S_{int}}$.
The constant $q_{D}$ collects the cell wall thickness, the diffusion
coefficient and the cell membrane surface area. The expression in
parentheses models the concentration gradient across the cell wall.
If the external analyte amount $\mathsf{S_{ext}}$ is sufficiently
high or subject to a constant supply, corresponding to \emph{Scenario
1 }described in \prettyref{sec:Bacterial-Sensor-Models}, it can be
treated as constant $\mathsf{\widetilde{S}_{ext}}$.

Considering that the uptake process is linear and time-invariant (LTI),
it can thus be represented in form of an impulse response given by
\begin{equation}
h_{\mathsf{up}}(t)=\mathcal{F}^{-1}\left\{ \frac{\frac{q_{D}}{V_{\mathrm{e}}}}{j\omega+\frac{q_{D}}{V_{\mathrm{i}}}+\delta_{s}}\right\} =u(t)\frac{q_{D}}{V_{\mathrm{e}}}e^{-\left(\frac{q_{D}}{V_{\mathrm{i}}}+\delta_{s}\right)t}\label{eq:uptake_impulse_response}
\end{equation}
where $u(t)$ is the unit step function and $\mathcal{F}^{-1}\{\cdot\}$
denotes the inverse Fourier transform. 

\subsubsection*{Facilitated Diffusion and Active Transport Mechanisms}

In facilitated diffusion molecules are transported into the cell by
means of membrane proteins or permeases. Facilitated diffusion is
a passive process and is driven by a concentration gradient. Active
transport mechanisms generally require an energy expenditure. As the
number of transport proteins on the cell surface is finite, both processes
are inherently saturable regardless of the external concentration
which is the main difference to free diffusion. 

With the signal or effector molecule $\mathsf{S_{ext}}$ and the surface
permease $\mathsf{M}$, the reaction equation for this process is
given by 
\begin{equation}
\ce{\mathsf{S_{ext}} + \mathsf{M} <=>[k_{\mathrm{b}}][k_{-\mathrm{b}}] [\mathsf{S_{ext}}\mathsf{M}] ->[k_{\mathrm{int}}]\mathsf{M} + \mathsf{S_{int}}}\label{eq:internalisation_react_eq}
\end{equation}
where $k_{\mathrm{b}}$ ($k_{-\mathrm{b}}$) is the forward (backward)
reaction rate constant of the binding reaction and $k_{\mathrm{int}}$
is the internalization rate constant. 

Mathematically, the molecular uptake mechanisms follow saturation
dynamics and can thus be described using Michaelis-Menten kinetics
as 
\begin{align}
\frac{\mathrm{d}\mathsf{S_{int}}}{\mathrm{d}t}^{+} & =\kappa_{\mathrm{up}}\left(\frac{\mathsf{S_{ext}}}{K_{\mathrm{up}}+\mathsf{S_{ext}}}\right)\label{eq:Facilitated_transport_MM_rre}
\end{align}
where the parameters $\kappa_{\mathrm{up}}$ and $K_{\mathrm{up}}$
are the maximum internalization rate and concentration at half maximum,
respectively. They are explicitly derived from \eqref{eq:internalisation_react_eq}
as $\kappa_{\mathrm{up}}=\mathsf{M}_{0}k_{\mathrm{int}}$, where $\mathsf{M}_{0}$
is the initial permease amount at $t=0$ and $K_{\mathrm{up}}=\frac{k_{-\mathrm{b}}+k_{\mathrm{int}}}{k_{\mathrm{b}}}$,
following Michaelis-Menten derivations. Similar modeling approaches
have been discussed in the literature \cite{buttonNutrientUptakeMicroorganisms1998,leveauTfdKGeneProduct1998}.
For instance, in the literature $K_{\mathrm{up}}$ is occasionally
used to characterize the kinetic uptake properties of a specific permease
\cite{vandermeerIlluminatingDetectionChain2004}. Eq. \eqref{eq:Facilitated_transport_MM_rre}
is non-linear and it is thus not straightforward to derive a transfer
function or impulse response for it. However, for relatively low concentrations
$\mathsf{S_{ext}}\ll K_{\mathrm{up}}$, eq. \eqref{eq:Facilitated_transport_MM_rre}
can be treated as linear which facilitates derivation of the corresponding
LTI model which is given by
\begin{equation}
h_{\mathsf{up}}(t)=u(t)\frac{\kappa_{\mathrm{up}}}{K_{\mathrm{up}}}e^{-\delta_{s}t}.\label{eq:uptake_active_impulse_response}
\end{equation}

\subsection{Transcription Factor Activation and Analyte Modification \label{subsec:Transcription-Factor-Activation}}

Inside the bacterial sensor cell, the effector molecules either directly
act as transcription factor $\mathsf{TF}$ for a specific gene or
undergo a series of reactions and interactions with regulatory proteins
or response regulators $\mathsf{RR}$, to form the $\mathsf{TF}$
complexes. This is described by the reaction equation
\begin{align}
\ce{a\mathsf{S_{int}} + b\mathsf{RR}} & \ce{<=> \underset{\mathsf{TF}}{\underbrace{[\mathsf{S_{int}}^{a}\mathsf{RR}^{b}]}}}.\label{eq:intemediary_forming_reaction}
\end{align}
$a$ and $b$ denote the moieties of the species partaking in reaction
\eqref{eq:intemediary_forming_reaction}. Depending on the precise
reaction nature, the conceivable reaction dynamics range from linear
(in case of a monomer transcription factor) to highly non-linear (for
multimers). This is reflected by the reaction rate equations
\begin{align}
\frac{\mathrm{d}\mathsf{TF}}{\mathrm{d}t} & =k_{1}\left(\frac{\mathsf{S_{int}}}{V_{\mathrm{i}}}\right)^{a}\mathsf{RR}^{b}-\delta_{\mathsf{TF}}\mathsf{TF}\label{eq:monomerisation_rre}
\end{align}
with $\delta_{\mathsf{TF}}$ denoting the degradation rate constant
of $\mathsf{TF}$. In \eqref{eq:monomerisation_rre}, the division
of $\mathsf{S_{int}}$ by the bacterial cell volume $V_{\mathrm{i}}$
is due to the units of the bi-molecular rate constant $k_{1}\left(\si{\micro\meter\tothe{3a}\per\minute}\right)$
\cite{gustafssonCharacterizingDynamicsQuorumSensing2005}.

Due to the product of two state variables $\mathsf{S_{int}}$ and
$\mathsf{RR}$ in \eqref{eq:monomerisation_rre}, and the resulting
non-linearity, the derivation of a system function or impulse response
is challenging. With the assumptions of singular moieties on part
of $\mathsf{S_{int}}$, i.e., $a=1$ and of abundance of response
regulator with $\mathsf{RR}\gg\mathsf{S_{int}}$, it is, however,
feasible. In this case $\mathsf{RR}^{b}$ can be treated as a constant
parameter because the reaction with $\mathsf{S_{int}}$ does not substantially
change the concentration of $\mathsf{RR}^{b}$. Then, the impulse
response can be given by
\begin{equation}
h_{\mathsf{mod}}(t)=\mathcal{F}^{-1}\left\{ \frac{\frac{k_{1}}{V_{\mathrm{i}}}\mathsf{RR}^{b}}{j\omega+\delta_{\mathsf{TF}}}\right\} =u(t)\frac{k_{1}}{V_{\mathrm{i}}}\mathsf{RR}^{b}\text{e}^{-\delta_{\mathsf{T}}t}.\label{eq:TF_impulse_response}
\end{equation}
In case of a multimerization with $a>1$, the process is highly non-linear
and an impulse response can only be derived by employing more elaborate
techniques such as Volterra series analysis \cite{veleticMolecularCommunicationModel2019}.

As stated above, a signal transduction pathway can comprise an indirect
uptake mechanism where the analyte molecule is not entering the cell
directly, but binds to a surface permease leading to the modification
of the intracellular response regulator. This process is termed \emph{phosphorylation.} 

\subsubsection{Phosphorylation\label{subsec:Phosphorylation}}

A common mechanism by which intracellular $\mathsf{RR}$ is turned
on is \emph{phosphorylation}. When the target analyte extracellularly
binds to the receptor on the cell surface, a phosphate group is transferred
from the intracellular surface receptor component to the free $\mathsf{RR}$.
For an arbitrary $\mathsf{RR}$ (with phosphorylated form $\mathsf{RR}_{\mathrm{p}}$
which then acts as transcription factor) and a surface receptor $\mathsf{C}$
(with bound state $\mathsf{C_{b}}$) this can be modeled as \cite{gustafssonCharacterizingDynamicsQuorumSensing2005}
\begin{align}
\frac{\mathrm{d}\mathsf{C_{b}}}{\mathrm{d}t} & =k_{\mathrm{B}}\frac{\mathsf{S_{ext}}}{V_{\mathrm{e}}}\mathsf{C}-k_{\mathrm{B}_{-}}\mathsf{C_{b}}-\delta_{\mathsf{C_{\mathrm{b}}}}\mathsf{C_{b}}\label{eq:phosphorylation_receptor_binding}\\
\frac{\mathrm{d}\mathsf{RR}_{\mathrm{p}}}{\mathrm{d}t} & =k_{\mathrm{p}}\mathsf{RR}\mathsf{\frac{C_{b}}{V_{\mathrm{i}}}}-k_{\mathrm{p}_{-}}\mathsf{RR}_{\mathrm{p}}-\delta_{\mathsf{RR}_{\mathrm{p}}}\mathsf{RR}_{\mathrm{p}}.\label{eq:phosphorylation}
\end{align}
 $k_{\mathrm{B}}$, $k_{\mathrm{B}_{-}}$, $k_{\mathrm{p}}$ and $k_{\mathrm{p}_{-}}$
are the reaction rate constants for binding, unbinding, phosphorylation
and dephosphorylation, respectively. We note that the total number
of surface receptors in free or bound state is constant, i.e., $\widetilde{\mathsf{c}}=\mathsf{C}+\mathsf{C_{b}}$
following the \emph{conservation of mass }rule. Under the condition
that the phosphorylation of $\mathsf{RR}$ does not affect its stability
($\delta_{\mathsf{RR}}=\delta_{\mathsf{RR}_{\mathrm{p}}}$), it also
applies that $\widetilde{\mathsf{RR}}=\mathsf{RR}+\mathsf{RR}_{\mathrm{p}}$.
From this follows that \eqref{eq:phosphorylation_receptor_binding}
and \eqref{eq:phosphorylation} can be written without explicit dependence
on $\mathsf{C}$ and $\mathsf{RR}$ as 
\begin{align}
\frac{\mathrm{d}\mathsf{C_{b}}}{\mathrm{d}t} & =k_{\mathrm{B}}\mathsf{\frac{\mathsf{S_{ext}}}{V_{\mathrm{e}}}}\left(\widetilde{\mathsf{c}}-\mathsf{C_{b}}\right)-k_{\mathrm{B}_{-}}\mathsf{C_{b}}-\delta_{\mathsf{C_{\mathrm{b}}}}\mathsf{C_{b}}\label{eq:phosphorylation_receptor_binding_rewritten}\\
\frac{\mathrm{d}\mathsf{\mathsf{RR}_{\mathrm{p}}}}{\mathrm{d}t} & =k_{\mathrm{p}}\mathsf{\frac{C_{b}}{V_{\mathrm{i}}}}\left(\widetilde{\mathsf{RR}}-\mathsf{\mathsf{RR}_{\mathrm{p}}}\right)-k_{\mathrm{p}_{-}}\mathsf{\mathsf{RR}_{\mathrm{p}}}-\delta_{\mathsf{RR}_{\mathrm{p}}}\mathsf{\mathsf{\mathsf{RR}_{\mathrm{p}}}}.\label{eq:phosphorylation_rewritten}
\end{align}

\subsection{Transcription and Translation \label{subsec:Transcription-and-Translation}}

In the final step of the detection chain of a bacterial sensor, the
$\mathsf{TF}$ complex binds to DNA segments known as promoter and
thereby regulates gene expression. This regulation can manifest either
as up-regulation (activation) or repression (inhibition; not treated
here) of gene expression. The precise mechanism that underlies this
essential process justifies closer inspection which is provided in
Appendix \prettyref{sec:Modeling-Gene-Transcription}. This analysis
finally yields the reaction rate equation for a reporter protein $\mathsf{P}$
as

\begin{equation}
\frac{\mathrm{d}\mathsf{P}}{\mathrm{d}t}=\alpha_{\mathrm{T}}\left[\kappa_{\mathrm{B}}+\kappa_{\mathsf{P}}\left(\frac{\mathsf{TF}^{n_{\mathrm{Hill}}}}{K^{n_{\mathrm{Hill}}}+\mathsf{TF}^{n_{\mathrm{Hill}}}}\right)\right]-\delta_{\mathsf{P}}\mathsf{P}.\label{eq:Protein_Hill_Activation}
\end{equation}
$\alpha_{\mathrm{T}}$ denotes the translation efficiency, $\kappa_{\mathrm{B}}$
is the basal transcription rate, $\kappa_{\mathsf{P}}$ is the maximum
up-regulated transcription rate, $K$ is the Michaelis-Menten constant
for transcription, $n_{\mathrm{Hill}}$ is the Hill coefficient and
$\delta_{\textsf{P}}$ the decay rate constant of the reporter protein. 

With $\mathsf{TF}\ll K$, $n_{\mathrm{Hill}}=1$ and assuming no basal
transcription ($\kappa_{\mathrm{B}}=0$), eq. \eqref{eq:Protein_Hill_Activation}
can be modeled as an LTI system, i.e., 
\begin{equation}
h_{\mathsf{T}}(t)=u(t)\alpha_{\mathrm{T}}\frac{\kappa_{\mathsf{P}}}{K}e^{-\delta_{\mathsf{P}}t}.\label{eq:LTI_model_transcription}
\end{equation}

\section{Bacterial Sensor Models \label{sec:Bacterial-Sensor-Models}}

With the mathematical descriptions for the single steps in the detection
chain, it is now possible to construct full mathematical models of
different bacterial sensor architectures. First, we outline under
which conditions a bacterial sensor follows an LTI model. We then
briefly discuss how the effects of bacterial population dynamics can
be accounted for in a population-level bacterial sensor. Finally,
we present a full bacterial sensor model of a real bacterial sensing
system based on the strain \emph{Streptococcus mutans.} 

\subsection{Linear Model for Bacterial Sensor \label{subsec:Linear-Model-for}}

\begin{figure}
\centering{}\includegraphics[width=0.95\columnwidth,height=0.15\textheight,keepaspectratio]{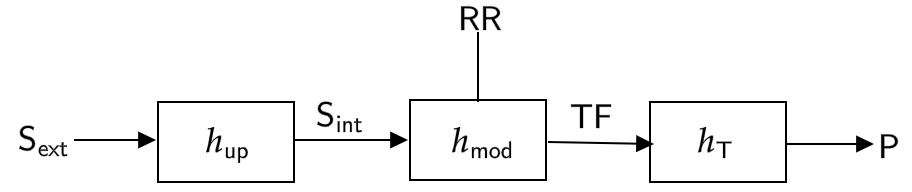}\caption{\label{fig:Block-diagram-of-bacterial-sensor}Block diagram of the
LTI model of a bacterial sensor.}
\end{figure}
It is possible to construct a \emph{linear} bacterial sensor model
if the following conditions are met:
\begin{itemize}
\item Uptake via diffusion \emph{or }active uptake modeled using Michaelis-Menten
kinetics with analyte concentrations $\mathsf{S_{ext}}\ll K_{\mathrm{up}}$
\item Analyte modification following \eqref{eq:TF_impulse_response} with
analyte moiety of $a=1$
\item Transcription and translation following a Michaelis-Menten model with
$n_{\mathrm{Hill}}=1$ with $\mathsf{TF}\ll K$
\end{itemize}
The full detection chain of the sensor is then described by an LTI
model visualized in \prettyref{fig:Block-diagram-of-bacterial-sensor}
where each step in the detection chain, i.e., uptake, modification
and transcription/translation is symbolized by a block representing
the process's impulse response. The significance of this approach
is that the output protein P can be expressed directly as
\begin{equation}
\mathsf{P}(t)=\underset{h_{\mathsf{BS}}(t)}{\underbrace{h_{\mathsf{up}}(t)*h_{\mathsf{mod}}(t,\mathsf{RR})*h_{\mathsf{T}}(t)}}*\mathsf{S_{ext}}(t)\label{eq:bacterial_sensor_linear_model}
\end{equation}
where $*$ denotes convolution and $h_{\mathsf{BS}}(t)$ the total
impulse response of the bacterial sensor cell. In \eqref{eq:bacterial_sensor_linear_model},
the parametrization of $h_{\mathsf{mod}}(t,\mathsf{RR})$ by $\mathsf{RR}$
is made explicit to emphasize the condition under which \eqref{eq:TF_impulse_response}
applies. 

\subsection{Whole Cell Bacterial Sensor - Population Dynamics}

A population-level bacterial sensor harvests the bioluminescent response
of an entire bacterial population; it is thus necessary to incorporate
the population size into the model. Consequently, $\nu(t)$ models
the size of the bacterial sensor strain\emph{ }at time $t$. 

Bacterial populations undergo specific phases in the course of their
growth cycle: the lag phase, the log-phase, the stationary phase and
the death phase. Numerous mathematical models and iterations on the
classical sigmoidal growth model for describing these dynamics have
been published \cite{zwieteringModelingBacterialGrowth1990}. Our
model is based on a variation of the original Verhulst growth model
\cite{pelegLogisticVerhulstModel2007}, where our contribution is
the incorporation of a \emph{delay }into the the governing differential
equation.

\begin{figure}[t]
\centering{}\includegraphics[width=1\columnwidth]{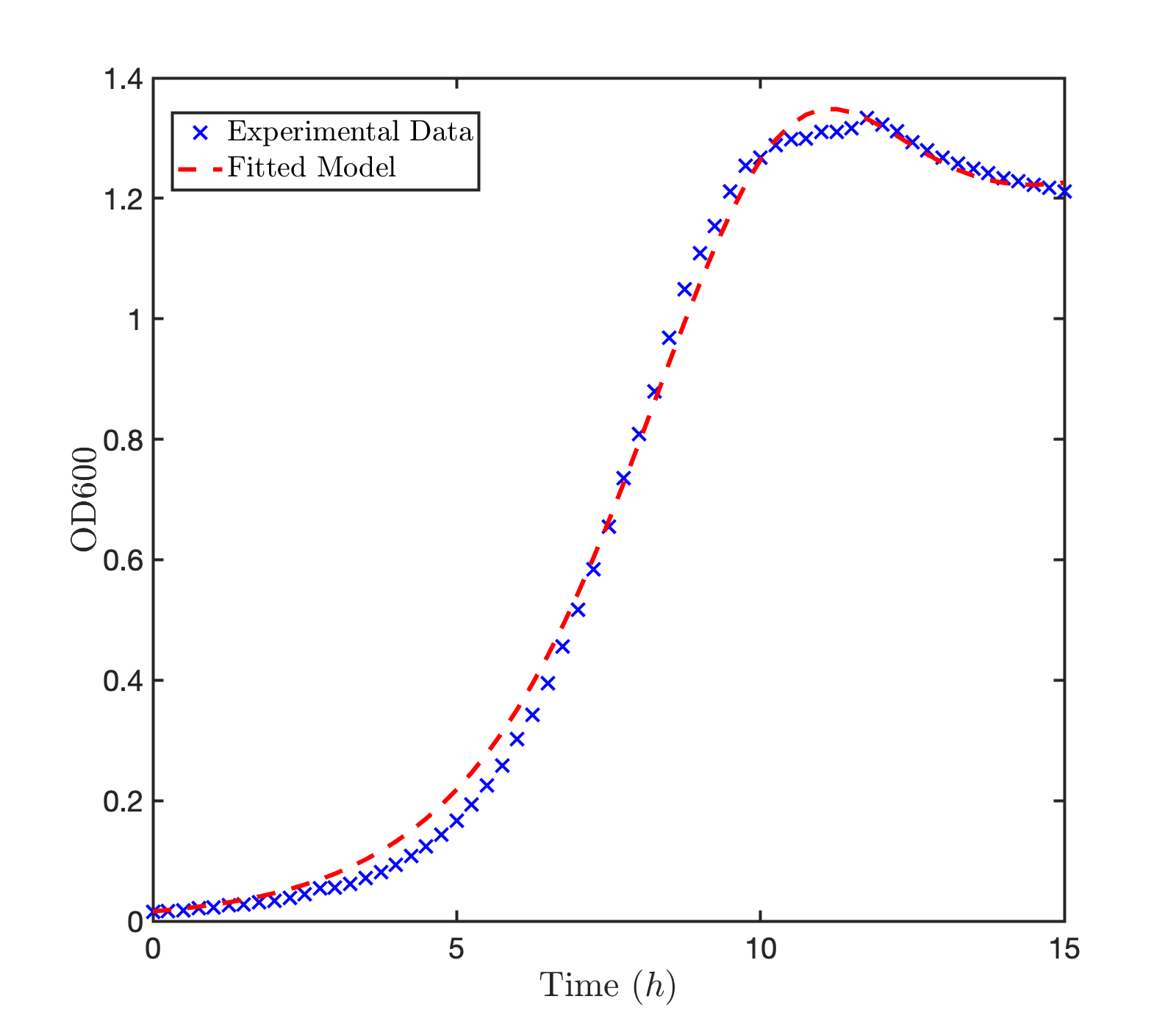}\caption{\label{fig:Experimental-bacterial-growth} Experimental bacterial
optical density (OD600) for \emph{Streptococcus mutans} vs. delay-differential
equation model with fitted parameters $K_{\nu}=7.5603\times10^{8}$,
$\mu=\SI{0.0090224}{\per\minute}$, $\tau=\SI{80.5438}{\minute}$
and $b_{\nu}=1$. }
\end{figure}

The sensor strain's growth dynamics are then more accurately described
by 
\begin{equation}
\frac{\mathrm{d}\nu}{\mathrm{d}t}=\mu\nu(t)\left(1-\frac{\nu(t-\tau)}{K_{\nu}}\right)^{b_{\nu}}\label{eq:bacterial_growth_model}
\end{equation}
where $\mu$ denotes the population growth constant and the carrying
capacity $K_{\nu}$ reflects the limits a bacterial population's growth
is subjected to in a specific environment. \textbf{$b_{\nu}$ }is
an additional scaling parameter that describes the populations capability
to exceed the carrying capacity ($b_{\nu}>1$) or to stay below it
($b_{\nu}<1$). Eq. \eqref{eq:bacterial_growth_model} is a delay-differential
equation with delay $\tau$ and is chosen to capture an overshoot
in the populations growth cycle where its size initially exceeds the
systems carrying capacity and subsequently levels off. We chose this
modeling approach to explain our experimental data obtained from wet-lab
experiments (\prettyref{fig:Experimental-bacterial-growth}). 

\subsection{Streptococcus mutans as Sensor for XIP\label{subsec:Streptococcus-mutans-as}}

For the bacterial sensor investigated in this paper, we choose the
bacterium \emph{Streptococcus mutans }(\emph{S. mutans})\emph{.} This
Gram-positive bacterial strain is found in the human oral cavity and
is the primary cause for tooth decay \cite{lemosBiologyStreptococcusMutans2019}.
In \emph{S. mutans}, natural competence, i.e., the ability of bacteria
to take up naked DNA from their environment and incorporate it into
their genomes as survival- and adaption strategy, is regulated by
pheromones/peptides \cite{kasparIntercellularCommunicationComXInducing2017}.
\emph{S. mutans }exhibits two signaling pathways over which two distinct
peptides function as signals: the CSP and the XIP pathway. Both pathways
activate the promoter of \emph{sigX,} the master regulator of competence.

\begin{figure}
\centering{}\includegraphics[width=0.95\columnwidth,height=0.35\textheight,keepaspectratio]{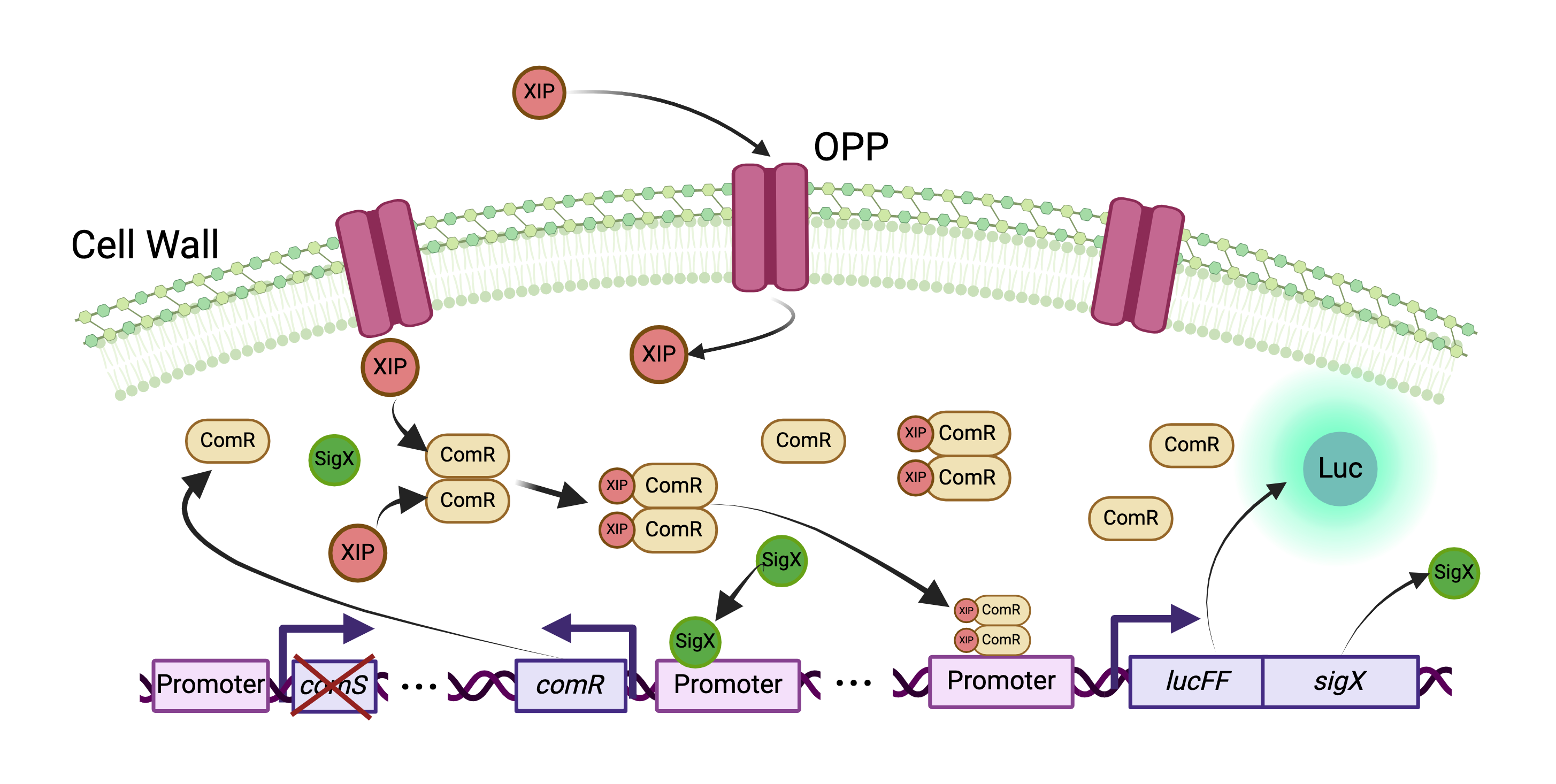}\caption{\label{fig:Sm_XIP_Signaling_Pathway}Transcriptional activation of
\emph{sigX/lucFF} operon in engineered \emph{S. mutans }strain\emph{.
}(Illustration was created with \emph{BioRender.com}).}
\end{figure}

In this paper, we focus on the distinct XIP signaling pathway. Extracellular
XIP (SigX-inducing peptide) is imported by the cell-membrane bound
oligopeptide permease ($\mathsf{OPP}$). The imported XIP activates
the free form of the intracellular response regulator ($\mathsf{\mathsf{ComR_{f}}})$
by building a dimer (i.e., involving two molecules of XIP and free
ComR each) $\mathsf{ComR_{m}}$ which, in turn, activates the transcription
of \emph{sigX.} Recent studies suggest that $\mathsf{ComR_{m}}$ further
acts as transcription factor for \emph{comR }\cite{khanPositiveFeedbackLoop2017}
which implies a positive feedback loop in the XIP signaling pathway
(\prettyref{fig:Sm_XIP_Signaling_Pathway}). In a wild-type \emph{S.
mutans }strain, $\mathsf{ComR_{m}}$ further acts as a transcription
factor for \emph{comS} which codes for the intracellular XIP-precursor
ComS. Since \emph{S. mutans }produces and secretes XIP (or ComS) naturally,
we knock out the \emph{comS }gene in the \emph{S. mutans }sensor strain
to ensure that no endogenous XIP is produced thereby eliminating this
source of `noise'.

For this experiment we use the \emph{lucFF }gene as a reporter system.
To construct the reporter system, first the \emph{sigX} promoter region
is amplified and cloned upstream of the luc gene. Both genes are consequently
expressed at the same rate when activated by the mature transcription
factor (\prettyref{fig:Sm_XIP_Signaling_Pathway}).

Complying with the steps in the detection chain of \emph{S. mutans},
we formulate \eqref{eq:XIP_int}-\eqref{eq:SigX} as a computational
model for the \emph{S. mutans}-XIP bacterial sensor. Eqs. \eqref{eq:XIP_int}
and \eqref{eq:XIP_ext} model the time evolution of the intracellular
XIP amount and extracellular XIP amount, respectively. The uptake
mechanism is modeled using an active uptake model. Eq. \eqref{eq:comRf}
captures the kinetics of the intracellular $\mathsf{ComR_{f}}$ regulator
which is transcribed at a basal rate $\kappa_{\mathrm{B}}$. $\mathsf{ComR_{f}}$'s
rate of production is further up-regulated by the master transcription
factor $\mathsf{SigX}$ which is reflected by the Michaelis-Menten
term in \eqref{eq:comRf} \cite{khanPositiveFeedbackLoop2017}. As
stated above, the immature $\mathsf{ComR_{f}}$ is activated by intracellular
$\mathsf{XIP_{int}}$ with which it forms a dimer which is the mature
form of $\mathsf{ComR_{f}}$ ($\mathsf{ComR_{m}}$) that acts as TF
for \emph{sigX.} The study on the ComRS pathway in \emph{Streptococcus
termophilus }\cite{haustenneModelingComRSSignaling2015} suggests
that the stoichiometry of the dimerization is such that \emph{two
ComR }and\emph{ two XIP }molecules are needed to produce \emph{one}
mature $\mathsf{ComR_{m}}$ molecule. This is reflected by the exponent
$2$\footnote{It must be noted that, in general, this exponent can also assume values
different to the moieties of the reactants. } in \eqref{eq:XIP_int}, \eqref{eq:comRf} and \eqref{eq:comRm},
and the factor $2$ in \eqref{eq:XIP_int} and \eqref{eq:comRf}.
Eqs. \eqref{eq:Luc} and \eqref{eq:SigX} model the gene transcription
and subsequent translation (reflected by the translation efficient
parameter $\alpha_{\mathsf{sigX}}$) of mRNA into the respective proteins,
luciferase and $\mathsf{SigX}$. The identical terms in both equations
reflect that luciferase and $\mathsf{SigX}$ are produced at \emph{identical
rate. }The gene expression of both genes \emph{sigX }and \emph{lucFF}
is activated by the transcription factor $\mathsf{ComR_{m}}$ and
the corresponding transcription rate is governed by Michaelis-Menten
kinetics. 

As stated, the total bioluminescence response $L_{\mathrm{B}}$ of
the sensor population is directly proportional to the amount of luciferase
in the system and is thus commonly modeled as 
\begin{equation}
L_{\mathrm{B}}=\xi_{\mathsf{lum}}\mathsf{Luc}\label{eq:lum_response_equation}
\end{equation}
where $\xi_{\mathsf{lum}}$ is a proportionality constant which collects
the effects of, for example, oxygen availability, substrate concentration,
etc. The model is highly simplifying and a more precise model (e.g.,
\cite{iqbalReconstructingPromoterActivity2017}) would additionally
consider residual substrates which participate in the bioluminescence
reaction. 
\begin{figure*}
\hrulefill{}

\begin{align}
\frac{\mathrm{d}\mathsf{XIP_{int}}}{\mathrm{d}t} & =\nu(t)\kappa_{\mathrm{up}}\left(\frac{\mathsf{XIP_{ext}}}{K_{\mathrm{up}}+\mathsf{XIP_{ext}}}\right)-\delta_{\mathsf{X_{int}}}\mathsf{XIP_{int}}\nonumber \\
\mathsf{} & -2k_{\mathsf{TF_{m}}}\left(\frac{1}{\nu(t)V_{\mathrm{i}}}\mathsf{XIP_{int}}\mathsf{ComR_{f}}\right)^{2}+2k_{\mathsf{TF_{f}}}\mathsf{ComR_{m}}\label{eq:XIP_int}\\
\frac{\mathrm{\mathrm{d}}\mathsf{XIP_{ext}}}{\mathrm{d}t} & =-\nu(t)\kappa_{\mathrm{up}}\left(\frac{\mathsf{XIP_{ext}}}{K_{\mathrm{up}}+\mathsf{XIP_{ext}}}\right)-\delta_{\mathsf{X_{ext}}}\mathsf{XIP_{ext}}\label{eq:XIP_ext}\\
\frac{\mathrm{d}\mathsf{ComR_{f}}}{\mathrm{d}t} & =\nu(t)\alpha_{\mathsf{ComR}}\kappa_{\mathrm{B}}-\delta_{\mathsf{TF}_{\mathrm{f}}}\mathsf{ComR_{f}}-2k_{\mathsf{TF}_{\mathrm{m}}}\left(\frac{1}{\nu(t)V_{\mathrm{i}}}\mathsf{XIP_{int}}\mathsf{ComR_{f}}\right)^{2}\nonumber \\
 & +2k_{\mathsf{TF}_{\mathrm{f}}}\mathsf{ComR_{m}}+\nu(t)\alpha_{\mathsf{comR}}\kappa_{\mathsf{x}}\left(\frac{\left(\frac{\mathsf{SigX}}{\nu(t)V_{\mathrm{i}}}\right)^{m}}{K_{\mathsf{Sig_{X}}}^{m}+\left(\frac{\mathsf{SigX}}{\nu(t)V_{\mathrm{i}}}\right)^{m}}\right)\label{eq:comRf}\\
\frac{\mathrm{d}\mathsf{ComR_{m}}}{\mathrm{d}t} & =k_{\mathsf{TF}_{\mathrm{m}}}\left(\frac{1}{\nu(t)V_{\mathrm{i}}}\mathsf{XIP_{int}}\mathsf{ComR_{f}}\right)^{2}\nonumber \\
 & -k_{\mathsf{TF}_{\mathrm{f}}}\mathsf{ComR_{m}}-\delta_{\mathsf{TF}_{\mathrm{m}}}\mathsf{ComR_{m}}\label{eq:comRm}\\
\frac{\mathrm{d}\mathsf{Luc}}{\mathrm{d}t} & =\nu(t)\alpha_{\mathsf{sigX}}\kappa_{\mathsf{comR}}\left(\frac{\left(\frac{\mathsf{ComR_{m}}}{\nu(t)V_{\mathrm{i}}}\right)^{n}}{K_{\mathsf{comR}}^{n}+\left(\frac{\mathsf{ComR_{m}}}{\nu(t)V_{\mathrm{i}}}\right)^{n}}\right)-\delta_{\mathrm{L}}\mathsf{Luc}\label{eq:Luc}\\
\frac{\mathrm{d}\mathsf{SigX}}{\mathrm{d}t} & =\nu(t)\alpha_{\mathsf{sigX}}\kappa_{\mathsf{comR}}\left(\frac{\left(\frac{\mathsf{ComR_{m}}}{\nu(t)V_{\mathrm{i}}}\right)^{n}}{K_{\mathsf{comR}}^{n}+\left(\frac{\mathsf{ComR_{m}}}{\nu(t)V_{\mathrm{i}}}\right)^{n}}\right)-\delta_{\mathrm{x}}\mathsf{SigX}\label{eq:SigX}
\end{align}

\hrulefill{}
\end{figure*}

\subsubsection{Parameter Estimation}

\begin{figure}
\centering{}\includegraphics[width=1\columnwidth]{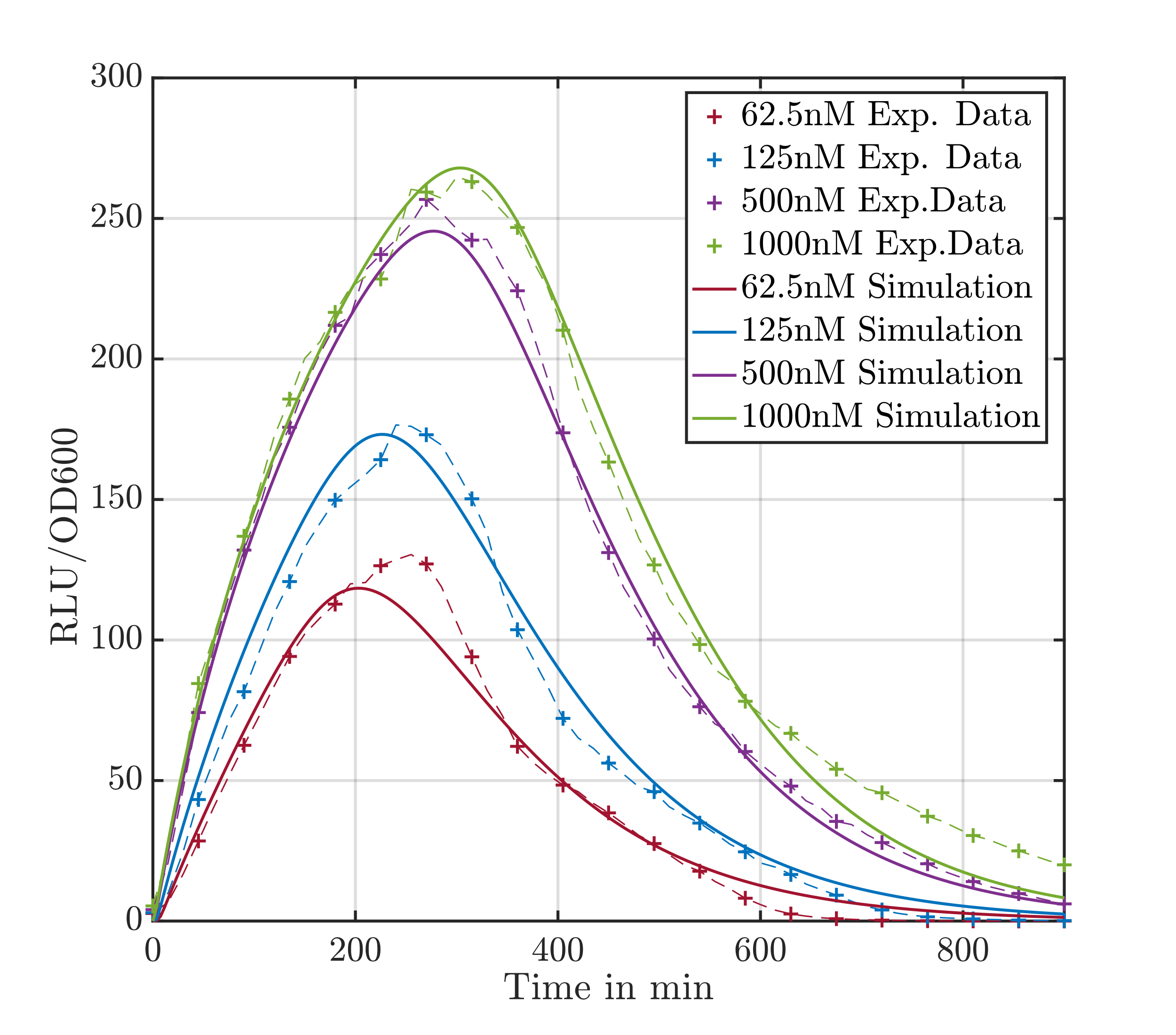}\caption{\label{fig:Experimental-vs.-simulation-parameter-fitting}Experimental
data vs. simulation of bacterial sensor luminescence response with
fitted parameters. Experimental data is RLU data normalized with experimental
OD600 bacterial growth curves.}
\end{figure}

We proceed to fit the computational \emph{S. mutans }model to experimentally
obtained data by adjusting the equation system's parameters. We consider
the \emph{average response }of a bacterial cell which is given by
\eqref{eq:XIP_int}-\ref{eq:SigX} \emph{discounting} the scaling
effect of the bacterial population factor $\nu(t)$. We use our wet-lab
data of the form $\text{RLU}/\text{OD600}$, i.e., luminescence data
in Relative-Light-Units (RLU) normalized with measured OD600 population
density curves (see \prettyref{fig:Experimental-bacterial-growth}).
The resulting data can be taken as measure of the average response
of a viable bacterial cell in the population. 

As we do not know $\xi_{\mathsf{lum}}$, estimating the parameters
of the model from the experimental data is greatly complicated. We
note that $\xi_{\mathsf{lum}}$ could experimentally be obtained by
calibrating the luminescence reader through parallel measurements
of the luciferase protein (e.g., western blotting) and comparison
with the obtained RLU curves. Here, we proceed by assuming a value
$\xi_{\mathsf{lum}}=10^{4}$ which is sensible in the context of reasonable
estimation ranges for the model parameters. We further possess only
limited observations of the system (in form of the measured luminescence
as proxy for the luciferase concentration), lacking data for the intermediary
species and processes. This, too, complicates estimating the full
system parameters as they cannot be verified against data observed
for the reactions they are \emph{directly }linked to. 

For the estimation, we restricted the optimization domain of parameters
to biologically sensible ranges. For several parameters we relied
on references in the literature as a guide for defining the estimation
ranges or to fix the parameter values in the estimation process \cite{haustenneModelingComRSSignaling2015}. 

The results of the parameter adjustment is presented in \prettyref{fig:Experimental-vs.-simulation-parameter-fitting}
and \prettyref{tab:XIP_System-Parameters}. The bacterial sensor system
has been implemented for simulation and parameter estimation in \emph{COPASI
}\cite{hoopsCOPASICOmplexPAthway2006}. The parameter estimation was
conducted using a \emph{Simulated Annealing }algorithm with a start
temperature of $T=1$ and a default cooling factor of $0.85$. For
our simulation and estimation we assumed a bacterial volume of $V_{\mathrm{i}}=\SI{1}{\micro\meter\tothe{3}}$
and an \emph{effective} extracellular volume of $V_{\mathrm{e}}=\SI{10}{\micro\meter\tothe{3}}$
per cell. 
\begin{table*}
\noindent \centering{}\caption{\emph{S. mutans}-XIP system parameters with adjusted values. \label{tab:XIP_System-Parameters} }
{\footnotesize{}}%
\begin{tabular}{>{\centering}p{1cm}>{\centering}p{1cm}c>{\raggedright}m{3cm}>{\centering}p{1cm}>{\centering}p{1cm}>{\centering}p{1cm}>{\raggedright}m{3cm}}
\hline 
\textbf{\footnotesize{}Parameter} & \textbf{\footnotesize{}Unit} & \textbf{\footnotesize{}Value} & \textbf{\footnotesize{}Description} & \textbf{\footnotesize{}Parameter} & \textbf{\footnotesize{}Unit} & \textbf{\footnotesize{}Value} & \textbf{\footnotesize{}Description}\tabularnewline
\hline 
\hline 
{\footnotesize{}$\mathsf{ComR_{f}}(0)$} & {\footnotesize{}-} & {\footnotesize{}$17.25$} & {\footnotesize{}Initial Response Regulator Amount} & {\footnotesize{}$\kappa_{\mathrm{B}}$} & {\footnotesize{}$\si{\per\minute}$} & {\footnotesize{}$0.492$} & {\footnotesize{}Basal comR Transcription Rate}\tabularnewline
\hline 
{\footnotesize{}$\kappa_{\mathrm{up}}$} & {\footnotesize{}$\si{\per\minute}$} & {\footnotesize{}$18.35$} & {\footnotesize{}Maximal Uptake Rate} & {\footnotesize{}$\kappa_{\mathsf{ComR}}$} & {\footnotesize{}$\si{\per\minute}$} & {\footnotesize{}$6.63$} & {\footnotesize{}Maximum ComR-induced sigX Transcription Rate}\tabularnewline
\hline 
{\footnotesize{}$K_{\mathrm{up}}$} & {\footnotesize{}$\si{\per\Vol}$} & {\footnotesize{}$333.6$} & {\footnotesize{}Michaelis-Menten Constant for Active Uptake} & {\footnotesize{}$K_{\mathsf{ComR}}$} & {\footnotesize{}$\si{\per\Vol}$} & {\footnotesize{}$14.02$} & {\footnotesize{}comR Michaelis-Menten Constant}\tabularnewline
\hline 
{\footnotesize{}$\delta_{\mathsf{XIP}}$} & {\footnotesize{}$\si{\per\minute}$} & {\footnotesize{}$0.023$} & {\footnotesize{}Degradation Constant of XIP} & {\footnotesize{}$\alpha_{\mathsf{SigX}}$} & {\footnotesize{}-} & {\footnotesize{}$2$} & {\footnotesize{}sigX Translation Efficiency}\tabularnewline
\hline 
{\footnotesize{}$k_{\mathsf{TF}_{\mathrm{m}}}$} & {\footnotesize{}$\si{\micro\meter\tothe{9}\per\minute}$} & {\footnotesize{}$0.02$} & {\footnotesize{}Transcription Factor Maturation Reaction Rate Constant} & {\footnotesize{}$\kappa_{\mathrm{X}}$} & {\footnotesize{}$\si{\per\minute}$} & {\footnotesize{}$0.216$} & {\footnotesize{}Maximum SigX induced comR Transcription Rate}\tabularnewline
\hline 
{\footnotesize{}$\delta_{\mathsf{TF}}$} & {\footnotesize{}$\si{\per\minute}$} & {\footnotesize{}$0.01$} & {\footnotesize{}Degradation Constant of ComR (fixed)\cite{haustenneModelingComRSSignaling2015}} & {\footnotesize{}$K_{\mathsf{SigX}}$} & {\footnotesize{}$\si{\per\Vol}$} & {\footnotesize{}$617413$} & {\footnotesize{}sigX Michaelis-Menten Constant}\tabularnewline
\hline 
{\footnotesize{}$k_{\mathsf{TF}_{\mathrm{f}}}$} & {\footnotesize{}$\si{\per\minute}$} & {\footnotesize{}$0.00001$} & {\footnotesize{}Transcription Factor De-Maturation Reaction Rate Constant} & {\footnotesize{}$\delta_{\mathrm{x}}$} & {\footnotesize{}$\si{\per\minute}$} & {\footnotesize{}$0.59$} & {\footnotesize{}SigX Degradation Constant}\tabularnewline
\hline 
{\footnotesize{}$\delta_{\mathsf{TF}_{\mathrm{m}}}$} & {\footnotesize{}$\si{\per\minute}$} & {\footnotesize{}$0.0027$} & {\footnotesize{}Degradation Constant of ComR-XIP Complex } & {\footnotesize{}$\delta_{\mathrm{L}}$} & {\footnotesize{}$\si{\per\minute}$} & {\footnotesize{}$0.0158$} & {\footnotesize{}Luciferase Degradation Constant (fixed)\cite{haustenneModelingComRSSignaling2015}}\tabularnewline
\hline 
{\footnotesize{}$\alpha_{\mathsf{ComR}}$} & {\footnotesize{}-} & {\footnotesize{}$0.166$} & {\footnotesize{}comR Translation Efficiency } & {\footnotesize{}$n$} & {\footnotesize{}-} & {\footnotesize{}$2.81$} & {\footnotesize{}Hill Coefficient}\tabularnewline
\hline 
{\footnotesize{}$m$} & {\footnotesize{}-} & {\footnotesize{}$3.96$} & {\footnotesize{}Hill Coefficient} &  &  &  & \tabularnewline
\hline 
\end{tabular}
\end{table*}

\subsubsection{Experimental procedure}

The experimental wet-lab data was obtained as follows: 

\paragraph{Bacterial strains and medium}

Cultures of \emph{S. mutans} (strain SM091 \cite{khanExtracellularIdentificationProcessed2012})
were grown in chemically defined medium (CDM) at $\SI{5}{\percent}$
$\ce{CO2}$, $\SI{37}{\degreeCelsius}$ and stored at $\SI{-80}{\degreeCelsius}$
in the same medium supplemented with $\SI{15}{\percent}$ glycerol
for cryopreservation. CDM is prepared as previously described \cite{wenderskaTranscriptionalProfilingOral2017}. 

\paragraph{Synthetic peptide}

The \emph{sigX}-inducing peptide (XIP) ($\ce{NH2-GLDWWSL-COOH}$;
$\SI{98}{\percent}$ purity; GenScript, Piscataway, NJ, USA) was reconstituted
with $\SI{20}{\micro\liter}$ dimethyl sulfoxide (DMSO) (Sigma-Aldrich,
St. Louis, MI, USA), to which $\SI{1}{\milli\liter}$ sterile distilled
water was added to give a final concentration of $\SI{10}{\milli\Molar}$
and stored at $\SI{-20}{\degreeCelsius}$. Working stocks were prepared
and aliquoted at $\SI{100}{\micro\Molar}$ in sterile distilled water
and stored at $\SI{-20}{\degreeCelsius}$. 

\paragraph{Luciferase reporter assays}

Culture stock of \emph{S. mutans} reporter strain at OD600 of $0.5$
was thawed and diluted $5$-fold in CDM. Serial dilutions of XIP were
made ($\SI{2000}{\nano\Molar}$, $\SI{1000}{\nano\Molar}$, $\SI{500}{\nano\Molar}$,
$\SI{250}{\nano\Molar}$, $\SI{125}{\nano\Molar}$, $\SI{62.5}{\nano\Molar}$,
$\SI{31.25}{\nano\Molar}$, $\SI{15.625}{\nano\Molar}$, $\SI{7.8}{\nano\Molar}$,
$\SI{3.9}{\nano\Molar}$, $\SI{1.9}{\nano\Molar}$ and control (No
XIP added) in $\SI{200}{\micro\liter}$ of CDM aliquoted in wells
of a $96$-well microtiter plate (Thermo Fisher Scientific). Then,
$\SI{100}{\micro\litre}$ of previously diluted \emph{S. mutans} culture
was added to the wells, in triplicates for each XIP concentration.
A volume of $\SI{10}{\micro\liter}$ of a $\SI{1}{\milli\Molar}$
aqueous D-luciferin solution (Synchem, Felsberg-Altenberg, Germany)
was added to each well, and three wells containing CDM and luciferin
without the inoculum were used as blanks. The plate was sealed with
a Top Seal (PerkinElmer), and incubated at $\SI{37}{\degreeCelsius}$
for $15$ hours. Relative luminescence units (RLU) and OD600 were
measured every $15$ minutes in a multi-detection micro plate reader
(Cytation 3; BioTek). 

\section{Engineering and Optimization of Bacterial Sensors \label{sec:Optimization-of-Bacterial-Sensors}}

While the focus so far was on the \emph{description} of a generic
bacterial sensor, we now proceed to investigate strategies for altering
and engineering a bacterial sensor with specific properties. It is
consequently useful to consider general aspects of sensor analysis
and engineering. Important sensor properties are 
\begin{itemize}
\item \textbf{Sensor sensitivity}: Differential response of the sensor to
different measurand quantities, i.e., the difference between sensor
outputs in response to different analyte concentrations.
\item \textbf{Response intensity}: Generally, a stronger response of the
bacterial sensor is clearly desirable for the maximization of the
Signal-to-Noise-Ratio (SNR). 
\item \textbf{Responsiveness:} Time between first contact with analyte and
\emph{final} sensor response, i.e., discounting contingent overshoots
during the transient response.\footnote{This is listed here for the sake of completeness but a formal treatment
of this sensor property is not within the scope of this work.}
\end{itemize}
In the following, we discuss how these general sensor properties (sensor
sensitivity and response intensity) can be evaluated in the context
of the investigated bacterial sensor models. As we aim at exploring
how bacterial sensors can be \emph{practically }altered and engineered
for specific purposes, it is necessary to establish a link between
\emph{parameters} in our computational models and \emph{biological
properties }that are accessible to manipulation in the wet-lab.

\subsection{Engineering of Bacteria - Linking Model Parameters to Biological
Properties}

Synthetic biology is devising ever more techniques for altering bacterial
dynamics for specific purposes, e.g., genetic or metabolic engineering.
For this paper, we are most interested in strategies that can directly
change a bacterial cell's properties that relate to the detection
chain discussed above. We therefore briefly review a number of techniques
by which bacteria can be manipulated to evoke desired sensor dynamics.
Here, we merely give a qualitative overview. In each specific case,
a quantitative relationship between a concrete manipulation measure
and the exact change in model parameter must be established which
is not within the scope of this work.

\paragraph*{Regulation of Uptake}

Active uptake mechanisms or phosphorylation-based internalization
are governed by cell membrane-bound ligands or channel molecules.
By regulating the expression of the genes coding for the specific
permease protein one has control of the uptake dynamics of a sensor.
Considering the \emph{S. mutans}-XIP\emph{ }sensor, by \emph{up-regulating
}the expression of the \emph{opp }gene we could \emph{increase} the
uptake rate constant $\kappa_{\mathrm{up}}$ due to its parametrization
by $\mathsf{M}_{0}$ which we introduced as the number of surface
permeases per cell above. 

\paragraph*{Regulation of Protein Stability}

The half-life of proteins involved in a biochemical system has a strong
impact on its overall dynamics. The degradation rates of proteins
in bacterial cell depend on a variety of factors: environmental conditions
(e.g., pH, temperature), the proteins structure and concentration,
or post-transcriptional modification (e.g., phosphorylation, bacterial
ubiquitination). Techniques based on\emph{ }attaching\emph{ degradation
tags }to proteins that mark them for degradation by the cellular protein
regulation mechanisms \cite{sekarNTerminalBasedTargetedInducible2016}
enhance the possibilities for engineering protein concentration level
in a targeted manner. Degradation tags can thus be used to directly
alter the protein degradation rate constants (see e.g., $\delta_{\mathsf{L}}$
in \eqref{eq:Luc}). 

\paragraph*{Regulation of Gene Expression}

Engineering of gene expression is the single most important tool in
synthetic biology. This comprises promoter engineering, plasmids which
can be engineered for specific purposes and inserted into bacteria,
and the targeted knocking out (see \emph{comS }in \emph{S. mutans}-XIP\emph{
}model above) or insertion of genes (e.g., promoter genes) directly
into the bacterial genome. The number of plasmids inserted into bacterial
cells, for example, directly modulates the expression levels of the
genes on the plasmid as more promoter sites are available for binding
transcription factors. According to Appendix \prettyref{sec:Modeling-Gene-Transcription},
the number of promoter sites thus directly modulates the transcription
rate (e.g., $\kappa_{\mathsf{comR}}$ in \eqref{eq:Luc}). Considering
the \emph{S. mutans-XIP }sensor example discussed above,\emph{ }this
can be used to regulate transcription of the \emph{sigX-lucFF }operon,
or to control $\kappa_{\mathrm{B}}$, i.e., the basal transcription
rate of the response regulator. This facilitates a finer regulation
of the modulation effect of the response regulator. 

\subsection{Steady-State Optimization }

\begin{figure}
\centering{}\includegraphics[width=1\columnwidth]{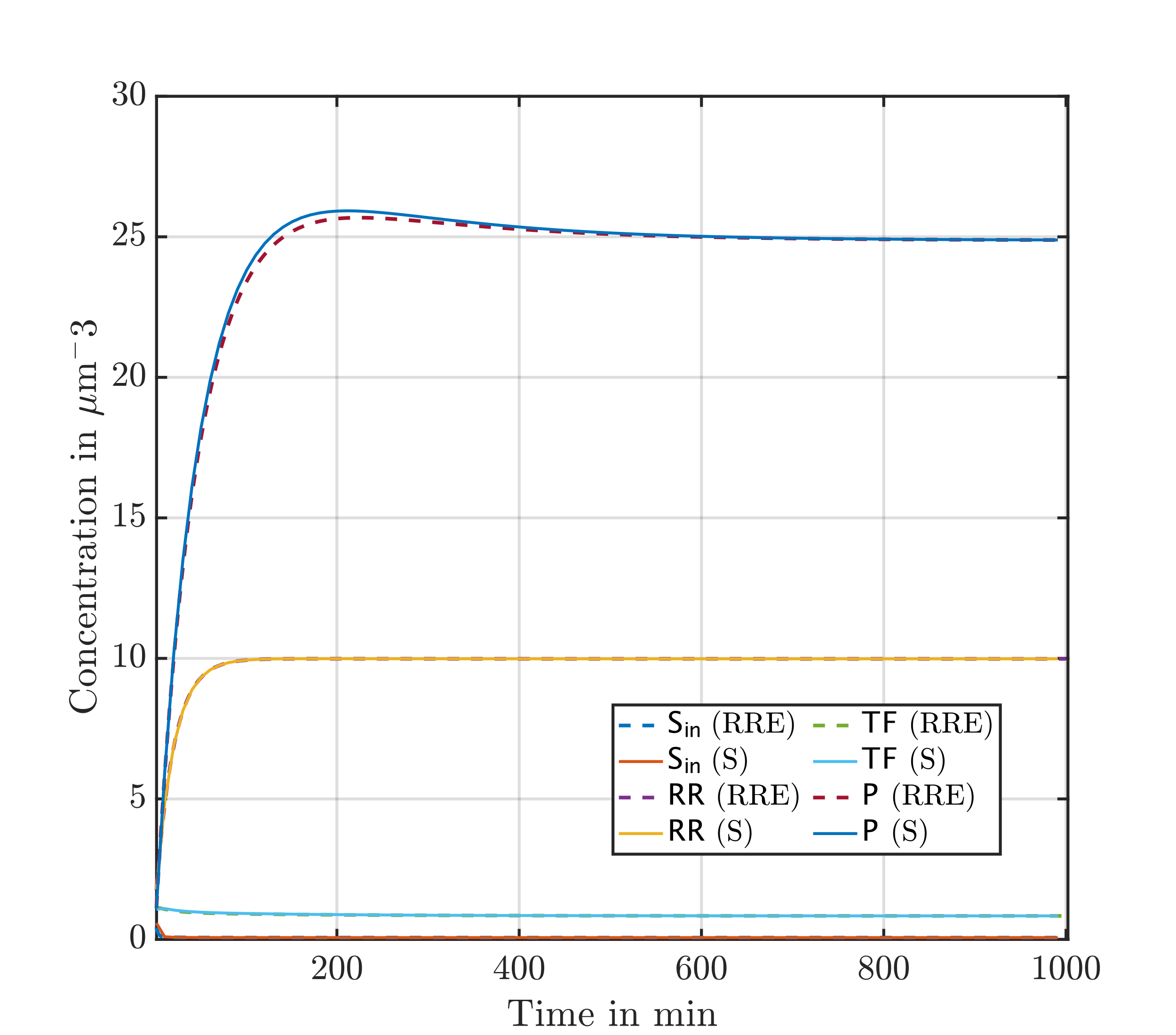}\caption{\label{fig:RRE--=000026-S-system-comparison}RRE- \& S-system model
numerical solution comparison. }
\end{figure}
Using the introduced S-system formalism, we construct the S-system
representation based on a RRE model of a generic bacterial sensor
following the same structure as the discussed \emph{S. mutans}-XIP
model and investigate it with respect to the sensor properties listed
above. This section treats the abundant-analyte case (corresponding
to \emph{Scenario 1 }introduced in \prettyref{sec:Bacteria-and-Biosensors}).
The parameter values of the RRE model used in this section are chosen
somewhat arbitrary without reference to an actual bacterial sensor
but we assumed realistic values for each parameter. For the transcription
factor maturation reaction we assumed moieties of $1$ for both reactants
but allowed for non-unity reaction orders $N$ and $M$. 

\subsubsection*{RRE Model}

\begin{align}
\frac{\mathrm{d}}{\mathrm{d}t}\mathsf{S_{in}} & =\kappa_{\mathrm{up}}\,\left(\frac{\mathsf{S_{ext}}}{K_{\mathrm{up}}+\mathsf{S_{ext}}}\right)+k_{\mathsf{TF}_{\mathrm{f}}}\,\mathsf{TF}\nonumber \\
 & -k_{\mathsf{TF}_{\mathrm{m}}}\,\mathsf{\mathrm{RR}}^{M}\,\mathsf{S_{\mathrm{in}}^{N}}-\delta_{\mathsf{S}}\,\mathsf{S_{\mathrm{in}}}\label{eq:S_in_GMA_eqs}\\
\frac{\mathrm{d}}{\mathrm{d}t}\mathrm{RR} & =\underset{\text{Basal Transcription}}{\alpha_{\mathsf{RR}}\underbrace{\kappa_{\mathsf{RR}}}}+k_{\mathsf{TF}_{\mathrm{f}}}\,\mathsf{TF}\nonumber \\
 & -k_{\mathsf{TF}_{\mathrm{m}}}\,\mathsf{RR}^{M}\,\mathsf{S}_{\mathsf{in}}^{N}-\delta_{\mathsf{RR}}\,\mathsf{RR}\label{eq:RR_GMA_eqs}\\
\frac{\mathrm{d}}{\mathrm{d}t}\mathrm{TF} & =k_{\mathsf{TF}_{\mathrm{m}}}\,\mathsf{RR}^{M}\,\mathsf{S}_{\mathsf{in}}^{N}\nonumber \\
 & -k_{\mathsf{TF}_{\mathrm{f}}}\,\mathsf{TF}-\delta_{\mathsf{TF}}\,\mathsf{TF}\label{eq:TF_gma_eqs}\\
\frac{\mathrm{d}}{\mathrm{d}t}\mathsf{P} & =\frac{\mathrm{\alpha}\,\kappa_{\mathrm{max}}\,\mathsf{TF}^{n_{\mathrm{Hill}}}}{K^{n_{\mathrm{Hill}}}+\mathsf{TF}^{n_{\mathrm{Hill}}}}-\delta_{\mathsf{P}}\,\mathsf{P}\label{eq:P_gma_eqs}
\end{align}

\subsubsection*{S-System Model}

\begin{align}
\frac{\mathrm{d}}{\mathrm{d}t}\mathsf{S}_{\mathsf{in},S} & =\alpha_{\mathsf{S}}\,\mathsf{TF}_{\mathsf{S}}^{g_{\mathsf{S,\mathsf{TF}}}}\,\mathsf{S}_{\mathsf{ext},S}^{g_{\mathsf{S_{ext,}S}}}\nonumber \\
 & -\beta_{\mathsf{S}}\,\mathsf{RR}_{S}^{h_{\mathsf{S,\mathsf{RR}}}}\,\mathsf{S}_{\mathsf{in},S}^{h_{\mathsf{S,S}}}\label{eq:S_in_S_eqs}\\
\frac{\mathrm{d}}{\mathrm{d}t}\mathsf{RR}_{S} & =\alpha_{\mathsf{RR}}\,\mathsf{TF}_{S}^{g_{\mathsf{RR},\mathsf{TF}}}\nonumber \\
 & -\beta_{\mathsf{RR}}\,\mathsf{RR}_{S}^{h_{\mathsf{RR,RR}}}\,\mathsf{S}_{\mathsf{in},S}^{h_{\mathsf{RR,S}}}\label{eq:R_S_eqs}\\
\frac{\mathrm{d}}{\mathrm{d}t}\mathsf{TF}_{S} & =\alpha_{\mathsf{TF}}\,\mathsf{RR}_{S}^{g_{\mathsf{TF,RR}}}\,\mathsf{S}_{\mathsf{in},S}^{g_{\mathsf{TF,S}}}\nonumber \\
 & -\beta_{\mathsf{\mathsf{TF}}}\,\mathsf{TF}_{S}^{h_{\mathsf{TF,TF}}}\label{eq:TF_S_eqs}\\
\frac{\mathrm{d}}{\mathrm{d}t}\mathsf{P}_{S} & =\alpha_{\mathsf{P}}\,\mathsf{RR}_{S}^{g_{\mathsf{P,RR}}}\,\mathsf{TF}_{S}^{g_{\mathsf{P,\mathsf{TF}}}}\nonumber \\
 & -\beta_{\mathsf{P}}\,\mathsf{P}_{S}^{h_{\mathsf{P,P}}}\label{eq:P_S_eqs}
\end{align}
\begin{figure*}[t]
\centering{}\subfloat[\label{fig:Influence_parameter_change_log_gain_sensor_strength}]{\centering{}\includegraphics[width=1\columnwidth]{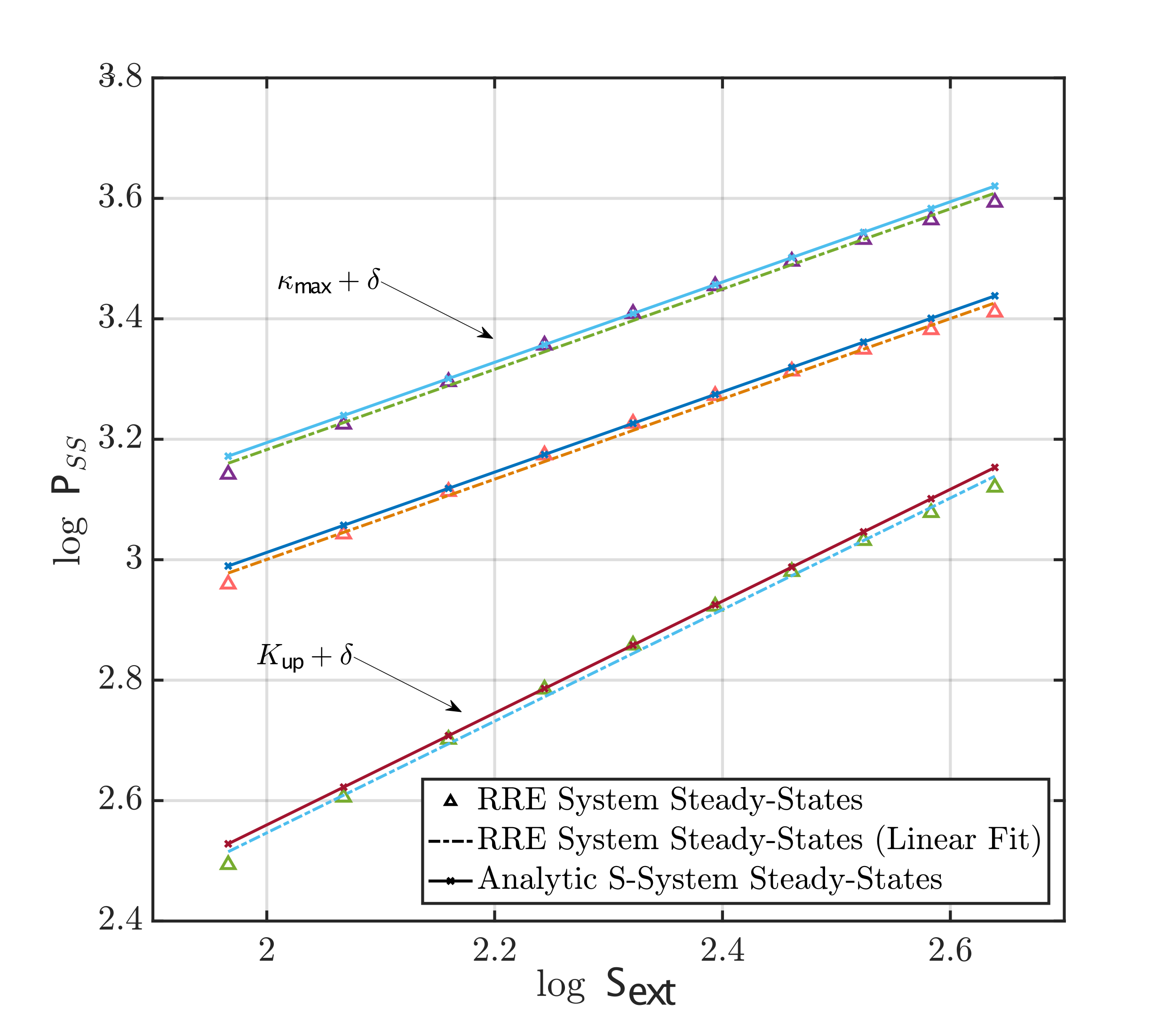}}\hfill{}\subfloat[\label{fig:Sensitivity-analysis-results}]{\centering{}\includegraphics[width=1\columnwidth]{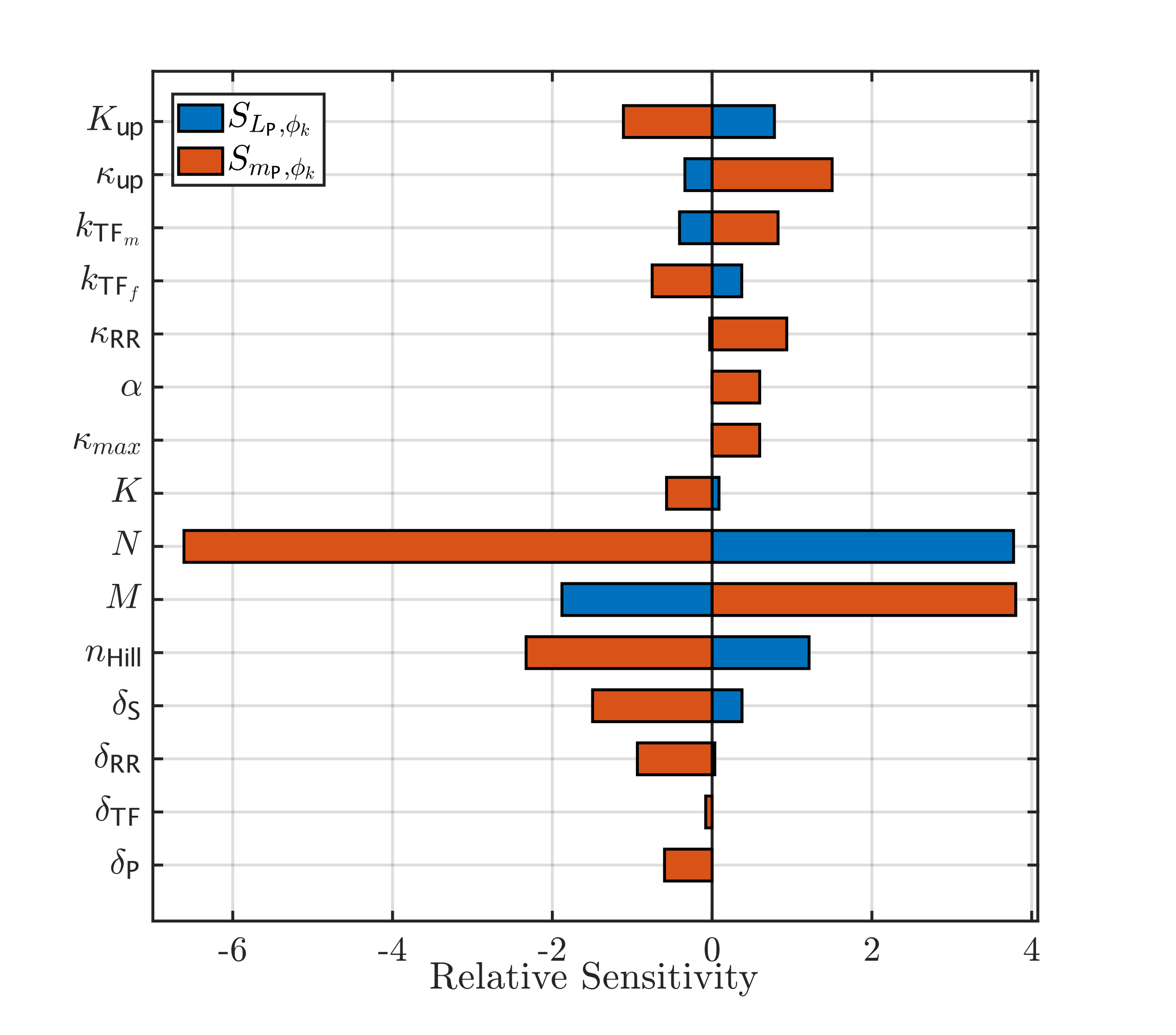}}\caption{\label{fig:sensitivity_analysis_results}S-system steady-state sensitivity
analysis. \textbf{(a)} Effect of parameter alterations on S-system
steady-state solutions. The center graph group represents the base
system with the initial parameter set.\textbf{ (b)} Sensitivities
with respect to model parameters.}
\end{figure*}

Eqs. \eqref{eq:S_in_GMA_eqs}-\eqref{eq:P_gma_eqs} are the RRE model
of the bacterial sensor and \eqref{eq:S_in_S_eqs}-\eqref{eq:P_S_eqs}
represent the corresponding S-system derived according to \eqref{eq:S_system_params_kinetic_rates}
and \eqref{eq:S_system_params_rate_consts}. The operating points
of the S-system representation $\mathbf{X}_{0}=\left[\mathsf{S}_{\mathsf{in},0},\mathsf{RR}_{0},\mathsf{TF}_{0},\mathsf{P}_{0}\right]^{T}$
are chosen so that the steady-state solutions of both systems are
near identical. The natural choice of $\mathbf{X}_{0}$ are the \emph{steady-state
solutions }of the RRE system. Solving the equation systems numerically
yields solution trajectories as plotted in \prettyref{fig:RRE--=000026-S-system-comparison}.
Note that the transient responses of the S- and RRE-system representations
can diverge significantly while the steady-state solutions are in
general identical. 

The S-system representation can then be analytically solved for steady-state
as in \eqref{eq:S_system_steady_state_eq} and the log-gain matrix
$\mathbf{L}$ obtained explicitly. Solving repeatedly for the steady-state
bacterial sensor output $\mathsf{P}_{SS}$ with different values of
the external/independent variable $\mathsf{S_{ext}}$ and plotting
in the log-log space yields a line whose slope reflects the log gain
of the sensor (\prettyref{fig:Influence_parameter_change_log_gain_sensor_strength}).
The S-system used to generate the data is constructed around the numerically
obtained RRE-system steady state solution for the mid-value of the
tested external variable $\mathsf{S_{ext}}$ range. This reflects
that a single S-system representation can accurately capture the system
dynamics responding to a wider range of input values. 

Next, the log-gain sensitivities $S_{L_{\mathsf{P}},\theta_{k}}$
and sensitivities $S_{m_{\mathsf{P}},\theta_{k}}$ are obtained using
\eqref{eq:log_gain_sensitivities} and \eqref{eq:m_sensitivities}
to measure the relative change in the log-gain and the constant component
of the steady-state sensor response in reaction to a relative change
in a RRE model parameter $\theta_{k}$. Importantly, and considering
the above mentioned sensor properties, $S_{L_{j},\theta_{k}}$ and
$S_{m_{j},\theta_{k}}$ constitute measures for \emph{sensor sensitivity
}and \emph{response} \emph{intensity},\emph{ }respectively.\emph{
}The results of this sensitivity analysis are shown in \prettyref{fig:Sensitivity-analysis-results}.

Noticeably, we find that $S_{L_{\mathsf{P}},\theta_{k}}$ and $S_{m_{\mathsf{P}},\theta_{k}}$
have opposite signs for a number of evaluated RRE parameters. This
suggests that changes in several RRE system parameters have \emph{diametric}
effects on \emph{sensor sensitivity }and \emph{response} \emph{intensity.
}In other words, according to this analysis \emph{increasing} the
steady-state response strength of a bacterial sensor by manipulating
a manipulable subset of these parameters (via techniques alluded to
above) \emph{diminishe}s its sensitivity and vice versa. 

\subsection{Dynamic Sensitivity Analysis }

While the preceding section investigates the \emph{steady-state} response
of a bacterial sensor, we now examine the case when the signal analyte
is scarce (i.e., non constant; see \emph{Scenario 2} described in
\prettyref{sec:Bacteria-and-Biosensors}) and either is depleted naturally
or enters the bacterial sensor system dynamics as a state variable
without a compensating influx. We are interested in how the findings
from the steady-state analysis map onto the time-course analysis of
the sensor response.

For the dynamic sensitivity analysis, $\mathsf{S_{ext}}$ cannot be
treated as external variable and enters the equation system \eqref{eq:S_in_GMA_eqs}-\ref{eq:P_gma_eqs}
as additional equation as (see also \eqref{eq:XIP_ext})
\begin{align}
\frac{\mathrm{d}}{\mathrm{d}t}\mathsf{S_{ext}}= & -\kappa_{\mathrm{up}}\,\left(\frac{\mathsf{S_{ext}}}{K_{\mathrm{up}}+\mathsf{S_{ext}}}\right)-\delta_{\mathsf{S}}\,\mathsf{S_{\mathrm{ext}}}.\label{eq:S_ext_eq_dyn_sens_analysis}
\end{align}
Here we work with the RRE system directly as our focus is on the transient
response. In particular, we isolate the \emph{maximum response }as
the point of interest as is common in the related literature. As
above, we investigate the influence of model parameters on both sensor
sensitivity and sensor signal intensity. 

\begin{figure*}
\centering{}\subfloat[\label{fig:Relative-Dynamic-Time-Course}]{\centering{}\includegraphics[width=1\columnwidth]{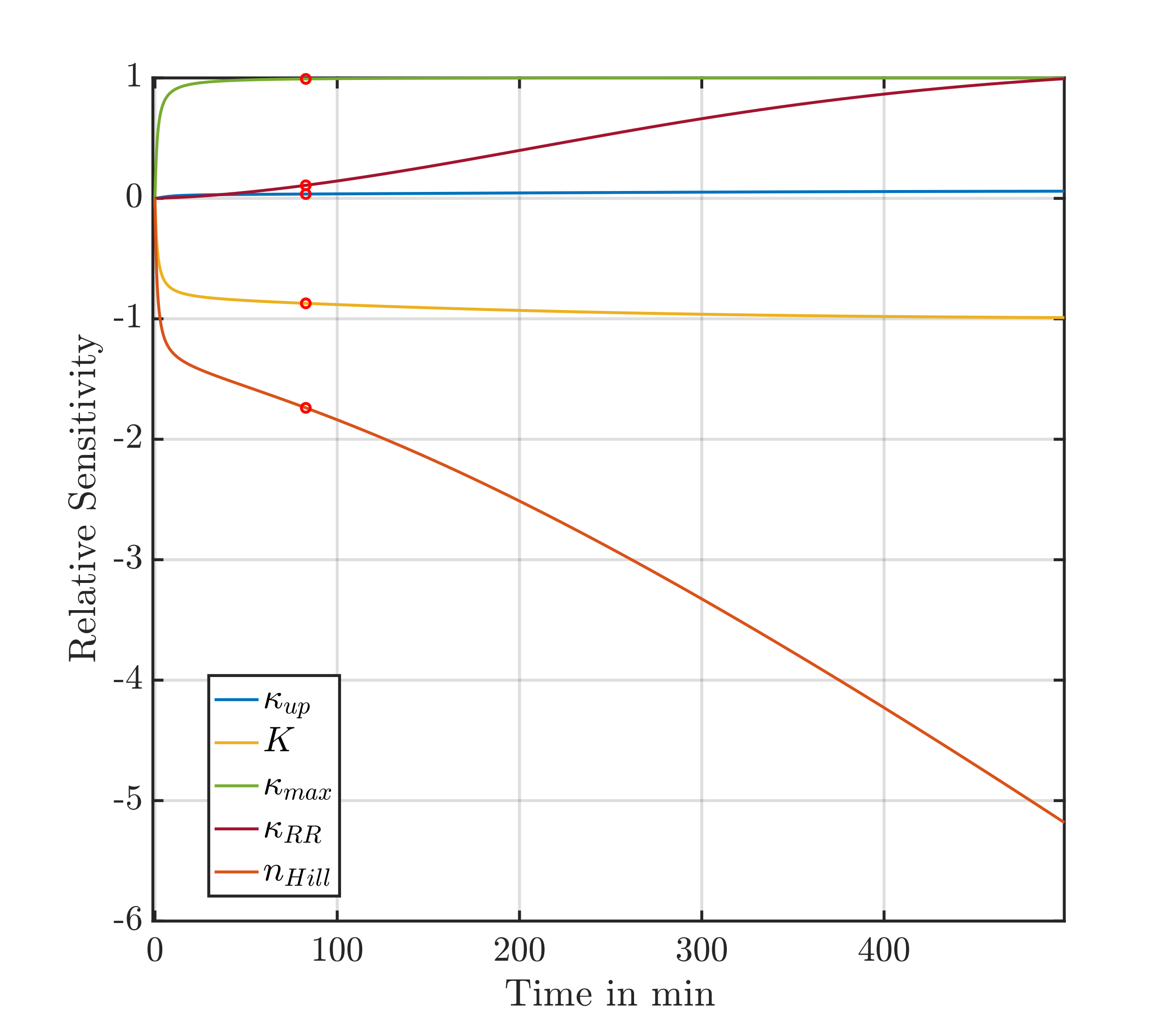}}\hfill{}\subfloat[\label{fig:Sensor-response-over}]{\centering{}\includegraphics[width=1\columnwidth]{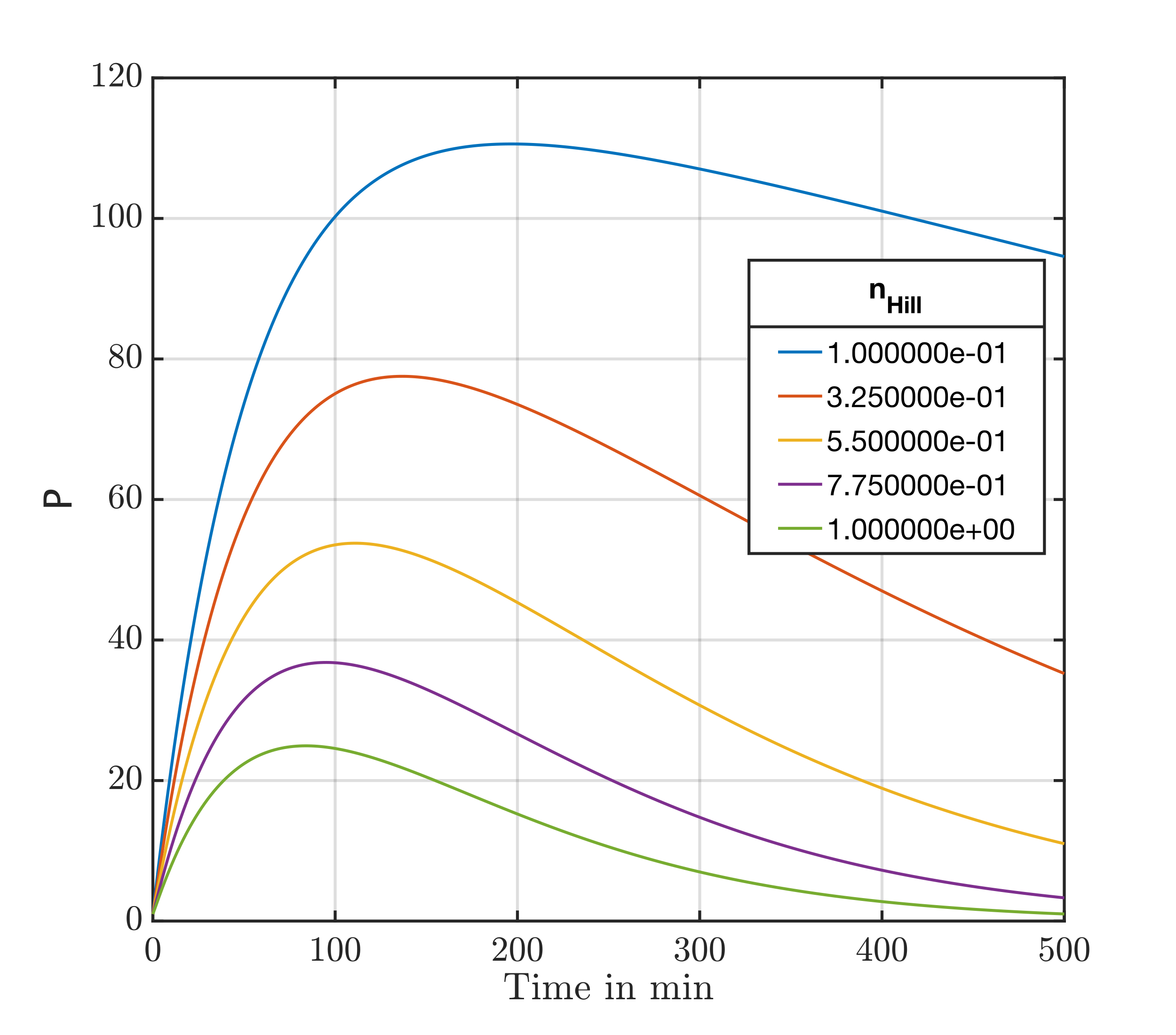}}\caption{\label{fig:Dynamic-Sensitivities}Dynamic Sensitivity Analysis. \textbf{(a)}
Relative Dynamic Time-Course Sensitivities. Red circles mark evaluation
at maximum response of $\mathsf{P}$. \textbf{(b)} Sensor response
over range of values of $n_{\mathrm{Hill}}$.}
\end{figure*}

Response intensity is investigated by appending one set of differential
equations \eqref{eq:dynamic_sensitivities_diff_eq} per investigated
parameter to the ODE system \eqref{eq:S_in_GMA_eqs}-\ref{eq:P_gma_eqs}
and solving the resulting system numerically with an appropriate solver
(we use the \noun{ode15s }solver in \noun{Matlab}). Thereby, trajectories
as in \prettyref{fig:Relative-Dynamic-Time-Course} are obtained which
are then sampled corresponding to the point of maximum transient response
of the sensor output $\mathsf{P}$, i.e., in the time-course solution
of the sensor response, the time of maximum response $\tau_{\max\mathsf{P}}$
is extracted and $S_{\mathsf{P},\theta_{k}}(\tau_{\max\mathsf{P}})$
evaluated. 

The entire dynamic sensitivity trajectory of the variable of interest
with respect to a specific parameter predicts how a change in the
parameter affects the time-course of the variable. For instance, the
dynamic sensitivity for the parameter $n_{\mathrm{Hill}}$ predicts
well how changes in it affect $\mathsf{P}$ (\prettyref{fig:Sensor-response-over}).

\begin{figure}
\centering{}\includegraphics[width=1\columnwidth]{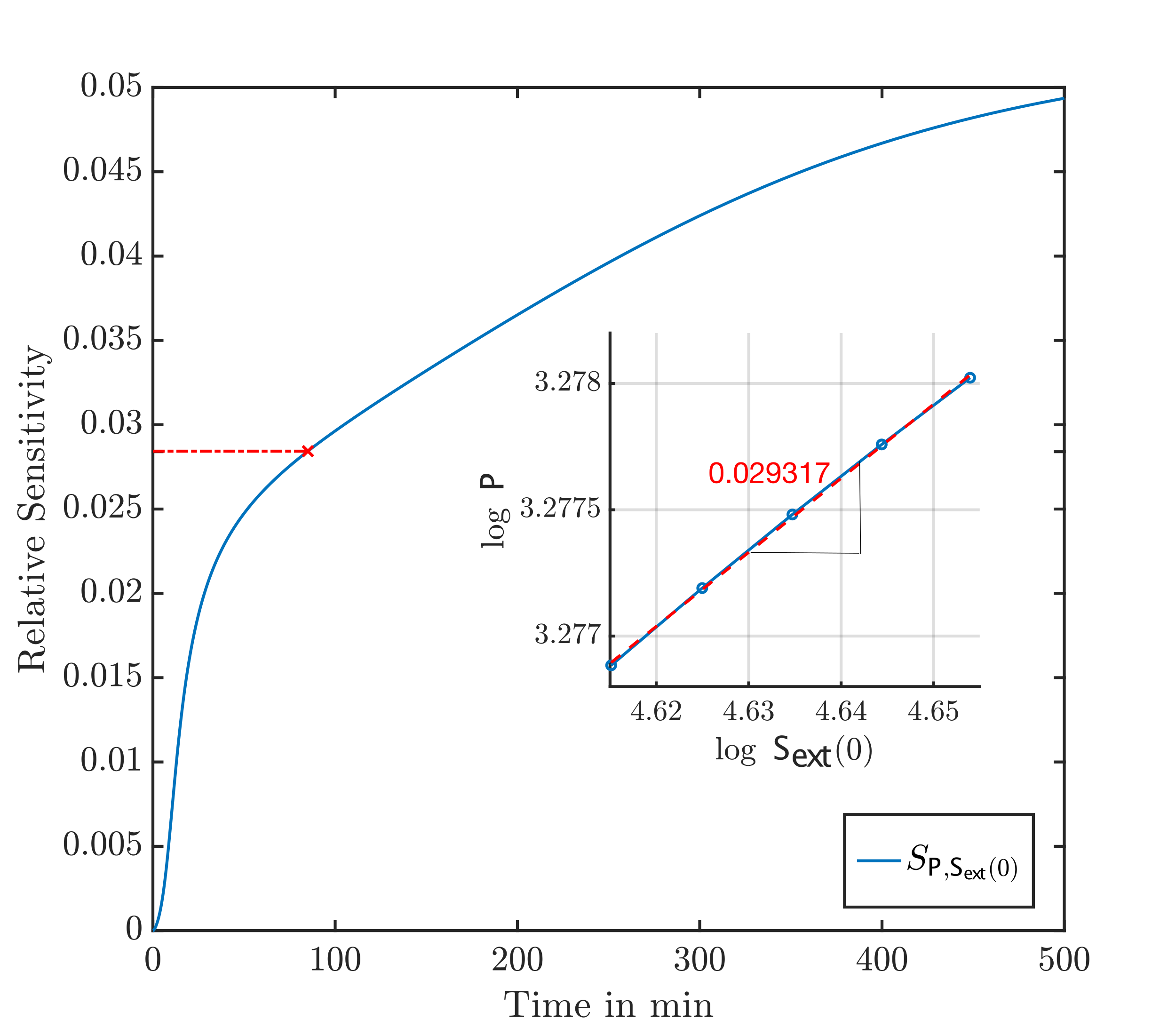}\caption{\label{fig:Dynamic-Sensitivity-Sext0}Dynamic Sensitivity with respect
to $\mathsf{S_{ext}}(0)$. }
\end{figure}

As described in \prettyref{subsec:Dynamics-Sensitivities}, we can
compute the dynamic sensitivity with respect to the amount of a species
(state-variable) at $t=0$ (i.e., the initial conditions of the corresponding
ODE system). Doing so for $\mathsf{P}$ with respect to $\mathsf{S_{ext}}(0)$,
i.e., calculating $S_{\mathsf{P},\mathsf{S_{ext}(0)}}=\frac{\partial\mathsf{P}}{\partial\mathsf{S_{ext}}(0)}$,
yields the results shown in \prettyref{fig:Dynamic-Sensitivity-Sext0}
where the sampled value, i.e., $S_{\mathsf{P},\mathsf{S_{ext}(0)}}(\tau_{\max\mathsf{P}})$,
corresponds well to the slope of the line in the inset plot. The latter
shows the maximum response of $\mathsf{P}$ evaluated for several
values of $\mathsf{S_{ext}}(0)$ in log-log space. 

\begin{figure*}
\centering{}\subfloat[\label{fig:dyn_param_var_effect}]{\centering{}\includegraphics[width=1\columnwidth]{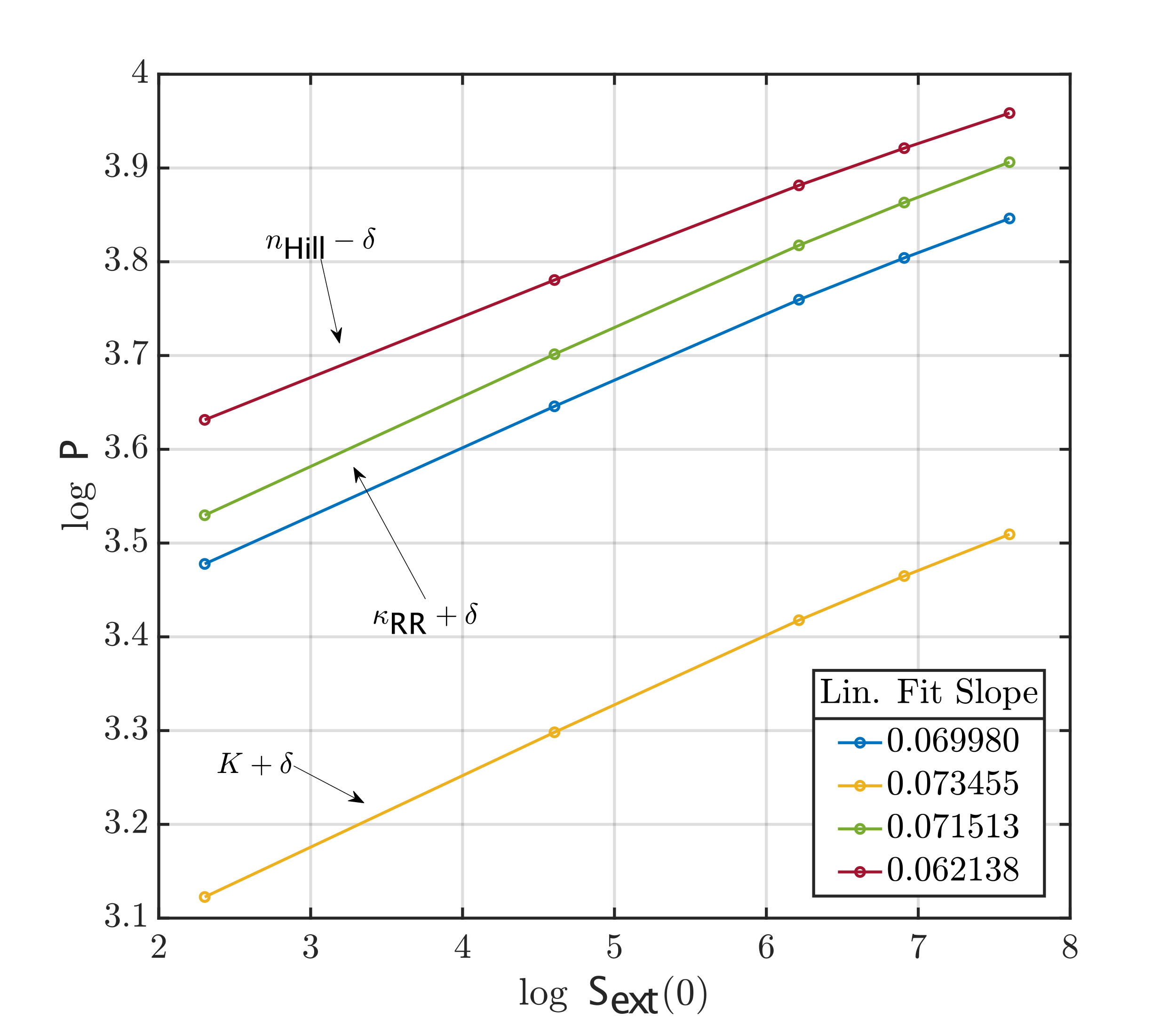}}~\subfloat[\label{fig:dyn_bar_plot}]{\centering{}\includegraphics[width=1\columnwidth]{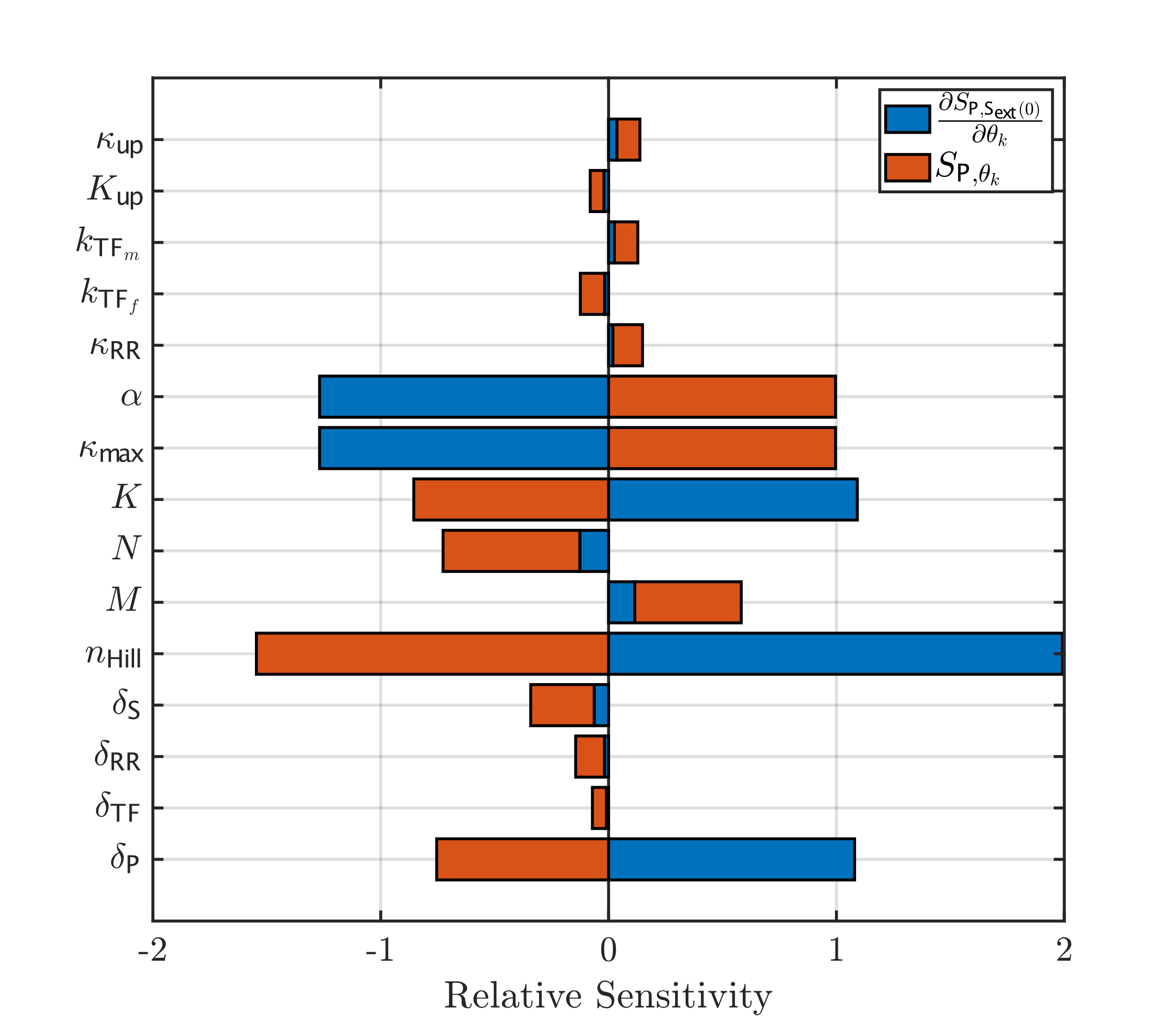}}\caption{\label{fig:Dynamic-Sensitivity-Analysis-Results}Dynamic Sensitivity
Analysis Results. \textbf{(a) }Effect of parameter alterations.\textbf{
}The blue line represents the base system with unaltered parameters.
\textbf{(b) }Sensitivities for model parameters.}
\end{figure*}

Equivalent to the steady-state log-gain sensitivities, we then proceed
to investigate the impact of model parameters on $S_{\mathsf{P},\mathsf{S_{ext}(0)}}$
by solving \eqref{eq:deq_dyn_log_gain_sensitivity} for all $\theta_{k}\in\mathbf{\Theta}_{R}$.
The corresponding results are presented in \prettyref{fig:Dynamic-Sensitivity-Analysis-Results}. 

Comparing \prettyref{fig:Sensitivity-analysis-results} and \prettyref{fig:dyn_bar_plot}
reveals that the sensitivities for steady-state and transient maximum
exhibit notable differences. While, for instance, evaluating $S_{L_{\mathsf{P}},\kappa_{\mathsf{RR}}}$,
$\frac{\partial S_{\mathsf{P},\mathsf{S_{ext}(0)}}}{\partial\kappa_{\mathsf{RR}}}$
and $S_{\mathsf{P},\mathsf{\kappa_{\mathsf{RR}}}}$ implies that a
positive change in $\kappa_{\mathsf{RR}}$ has only a minor effect
on the steady-state log-gain (namely \emph{decreasing }it), the maximum
transient response experiences both a positive shift in magnitude
\emph{and} a minor increase in sensitivity (as is also visualized
in \prettyref{fig:dyn_param_var_effect}). Other trends observed in
the stead-state analysis are, in contrast, reproduced by the dynamic
sensitivity analysis; alterations in $n_{\mathrm{Hill}}$, for example,
elicit strong diametric effects on signal intensity and sensitivity
for both cases. 

The sensitivity analysis results for both steady-state and transient
response presented in this section have validity in this form only
for the arbitrary parameter set alluded to above. However, the general
findings of the investigation could be confirmed for various tested
parameter sets. 

\section{Conclusion \label{sec:Conclusion}}

In this work we systematically investigated the quantitative properties
of the biochemical mechanisms governing bacterial sensor systems.
We dissected the bacterial sensor detection chain and provided an
overview of models for describing them. Based on this, we presented
a \emph{S. mutans}-XIP\emph{ }sensor as a case-example for which we
fitted the derived model to experimental wet-lab data. We obtained
a good match between our computational model and our empirical data
yielding biologically meaningful parameter values which are in the
expected ranges. While no general validity of the fitted parameter
values can be guaranteed, the results demonstrate the applicability
of the methodological approach. Our mathematical modeling approach
is limited to deterministic models, which is adequate for the scope
of this paper, but may be insufficient to study the inherent stochasticity
of molecular systems in greater detail. Further, we applied rather
restrictive assumptions (aqueous assays, homogeneous analyte distribution,
homogeneous expression levels of bacterial sensor cells) to the analysis.
These shortcomings may be addressed in future works. 

The main contribution of this paper is the analytic and numerical
derivation of common sensor-engineering measures, e.g., the \emph{sensor
sensitivity, }for bacterial sensors systems. This was achieved by
rigorous application of BST methods which yielded analytic results
for the steady-state treatment of bacterial sensors. We compared these
findings with results obtained from a dynamic time-course analysis
where we focused on the maximum response of a bacterial sensor. Specifically,
we mathematically investigated how various parameters from the computational
models influence log-gain (sensor sensitivity) and response strength
levels for both steady-state and maximum sensor response. We found
that, in many cases, with notable exceptions, parameter changes have
opposite effects on bacterial sensor sensitivity and response intensity
levels for both steady-state and maximum transient response. This
implies a trade-off which must be taken into account when designing
bacterial sensor systems. We mentioned the \emph{responsiveness}
of a sensor as crucial property but did not further investigate it
in this work; this might be addressed in a future study. 

We further identified and discussed a selection of parameters that
are most accessible for manipulation from a bio-engineering perspective.
A systematic investigation that links concrete manipulations in the
wet-lab to numerical changes in model parameters may be considered
for future research. 

By combining biological expertise with multifaceted computational
techniques, we aimed for complementing research in MC with a focus
on the quantitative aspects of engineering and optimizing bacterial
sensing systems. We expect that this work will contribute to guiding
cross-disciplinary researchers who want to computationally investigate
bacterial sensors. 

\appendices{}

\section{Modeling Gene Transcription \label{sec:Modeling-Gene-Transcription}}

For the sake of completeness, the dynamics of gene expression in a
bacterial sensor are summarized and a well-established mathematical
representation is derived (see e.g., \cite{unluturkGeneticallyEngineeredBacteriaBased2015a,dockeryMathematicalModelQuorum2001}):
\begin{itemize}
\item Transcription factor $\mathsf{TF}$\emph{ }binding to promoter $\mathsf{Pr}$
of a gene\emph{ }is reflected by the reaction equation
\begin{equation}
\ce{n\mathsf{TF} + \mathsf{Pr} <=>[k_{\mathrm{a}}][k_{\mathrm{d}}] \mathsf{TF}_{n} \ce{\mathsf{Pr}} }\label{eq:transcription_factor_promoter_binding-1}
\end{equation}
where $k_{\mathrm{a}}$ and $k_{\mathrm{d}}$ denote the forward and
backward reaction rate constants. Here, $n$ defines the affinity
of the binding process and denotes the number of transcription factor
molecules binding to a promoter site, thereby activating it and increasing
its affinity for binding polymerase and thus initiating transcription. 
\item Transcription of a gene into mRNA $\mathsf{p}_{\mathrm{mRNA}}$ follows
the reaction equation
\begin{equation}
\ce{\mathsf{TF}_{n} \mathsf{P} ->[k_{\mathrm{r}}] \mathsf{p}_{\mathrm{mRNA}} ->[\delta_{\mathrm{r}}] \emptyset}
\end{equation}
with transcription reaction rate constant $k_{\mathrm{r}}$ and decay
rate constant $\delta_{\mathrm{r}}$.
\item Translation of mRNA into protein $\textsf{P}$ with rate constant
$k_{\mathrm{T}}$ is reflected by
\begin{equation}
\ce{\mathsf{p}_{\mathrm{mRNA}} ->[k_{\mathrm{T}}] \mathsf{P}}
\end{equation}
\end{itemize}
It is practical to exploit the different time-scales of promoter-TF
binding and transcription, and the constant number of promoter sites
to derive a compact expression for the process. Consequently, the
transcription dynamics are conveniently described by Michaelis-Menten
or Hill kinetics where the former is a special case of the latter
and assumes a Hill coefficient of $n_{\mathrm{Hill}}=1$. $n_{\mathrm{Hill}}$
conceptually corresponds to the parameter in \eqref{eq:transcription_factor_promoter_binding-1}
where $n_{\mathrm{Hill}}$ can assume non-integer values. Consequently,
transcription can be modeled as,
\begin{equation}
\frac{\mathrm{d}\mathsf{p}_{\mathrm{mRNA}}}{\mathrm{d}t}=\kappa_{\mathsf{T}}\left(\frac{\mathsf{TF}^{n_{\mathrm{Hill}}}}{K^{n_{\mathrm{Hill}}}+\mathsf{TF}^{n_{\mathrm{Hill}}}}\right)-\delta_{\mathrm{r}}\mathsf{p}_{\mathrm{mRNA}}\label{eq:Transcription_Hill_Eq-1}
\end{equation}
with $K=\frac{k_{\mathrm{d}}+k_{\mathrm{r}}}{k_{\mathrm{a}}}$and
$\kappa_{\mathsf{T}}=\mathsf{Pr}_{0}k_{\mathrm{r}}$.

The mRNA is then translated into protein with translation rate constant
$k_{\mathrm{l}}$ following
\begin{equation}
\frac{\mathrm{d}\textsf{P}}{\mathrm{d}t}=k_{\mathrm{l}}\mathsf{p}_{\mathrm{mRNA}}-\delta_{P}\mathsf{P}.\label{eq:translation-1}
\end{equation}
With the quasi-steady-state assumption for $\mathsf{p}_{\mathrm{mRNA}}$
(based on mRNA being in general much shorter lived relative to the
corresponding protein which allows for the quasi-steady-state assumption
for $\mathsf{p}_{\mathrm{mRNA}}$), solving \eqref{eq:Transcription_Hill_Eq-1}
for $\mathsf{p}_{\mathrm{mRNA}}$ in steady state and inserting into
\eqref{eq:translation-1} gives
\begin{equation}
\frac{\mathrm{d}\mathsf{P}}{\mathrm{d}t}=\alpha_{\mathrm{T}}\kappa_{\mathsf{T}}\left(\frac{\mathsf{TF}^{n}}{K^{n}+\mathsf{TF}^{n}}\right)-\delta_{\mathsf{P}}\mathsf{P}\label{eq:Protein_MM-1}
\end{equation}
with $\alpha_{\mathrm{T}}=\frac{k_{l}}{\delta_{b}}$ as the \emph{translation
efficiency. }Eq. \eqref{eq:Protein_MM-1} can further include a constant
term $\kappa_{\mathrm{B}}$ which denotes the \emph{basal transcription
rate }at which DNA is transcribed in the absence of up-regulation.

\section{Derivation of Dynamic Log-Gain Sensitivity Equation\label{sec:Derivation-of-Dynamic}}

For the derivation of \eqref{eq:deq_dyn_log_gain_sensitivity} we
extend the argumentation resulting in \eqref{eq:dynamic_sensitivities_diff_eq}
\cite{dickinsonSensitivityAnalysisOrdinary1976}. Eqs. \eqref{eq:proof_dyn_log_gain_sens_start}
- \eqref{eq:proof_dyn_log_gain_sens} outline the derivation. In \eqref{eq:dlgsa_steps_explain}
the identity $\frac{\partial\dot{X}_{i}}{\partial X_{j}(0)}=0$ is
attributed to $X_{j}(0)$ not occurring explicitly in $\dot{X}_{i}$.
\begin{figure*}
\hrulefill{}

\begin{subequations}
\begin{align}
\frac{\mathrm{d}}{\mathrm{d}t}\left(\frac{\partial S_{i,X_{j}(0)}}{\partial\theta_{k}}\right) & =\frac{\mathrm{\partial}}{\mathrm{\partial}\theta_{k}}\left(\frac{\mathrm{d}S_{i,X_{j}(0)}}{\mathrm{d}t}\right)\label{eq:proof_dyn_log_gain_sens_start}\\
 & =\frac{\mathrm{\partial}}{\mathrm{\partial}\theta_{k}}\Biggl(\underset{=0}{\underbrace{\frac{\partial\dot{X}_{i}}{\partial X_{j}(0)}}}\frac{X_{j}(0)}{X_{i}}+\underset{=S_{i,X_{j}(0)}\frac{X_{i}}{X_{j}(0)}}{\underbrace{\frac{\partial X_{i}}{\partial X_{j}(0)}}}\frac{\mathrm{d}}{\mathrm{d}t}\left(\frac{X_{j}(0)}{X_{i}}\right)\Biggr)\label{eq:dlgsa_steps_explain}\\
 & =\frac{\mathrm{\partial}}{\mathrm{\partial}\theta_{k}}\left(\frac{\partial X_{i}}{\partial X_{j}(0)}\frac{\mathrm{d}}{\mathrm{d}t}\frac{X_{j}(0)}{X_{i}}\right)\\
 & =\frac{\mathrm{\partial}}{\mathrm{\partial}\theta_{k}}\left(S_{i,X_{j}(0)}\frac{X_{i}}{X_{j}(0)}\right)\frac{\mathrm{d}}{\mathrm{d}t}\left(\frac{X_{j}(0)}{X_{i}}\right)+\frac{\partial X_{i}}{\partial X_{j}(0)}\frac{\mathrm{\partial}}{\mathrm{\partial}\theta_{k}}\left(\frac{\mathrm{d}}{\mathrm{d}t}\left(\frac{X_{j}(0)}{X_{i}}\right)\right)\\
 & =\left(\frac{\partial S_{i,X_{j}(0)}}{\partial\theta_{k}}X_{i}+S_{i,X_{j}(0)}\frac{\partial X_{i}}{\partial\theta_{k}}\right)\frac{\mathrm{d}}{\mathrm{d}t}X_{i}^{-1}+X_{j}(0)\frac{\partial X_{i}}{\partial X_{j}(0)}\\
 & =-\left(\frac{\partial S_{i,X_{j}(0)}}{\partial\theta_{k}}X_{i}+S_{i,X_{j}(0)}\frac{\partial X_{i}}{\partial\theta_{k}}\right)\frac{\dot{X_{i}}}{X_{i}^{2}}+X_{j}(0)\frac{\partial X_{i}}{\partial X_{j}(0)}\left(2X_{i}^{-3}\dot{X_{i}}-X_{i}^{-2}\frac{\partial\dot{X_{i}}}{\partial\theta_{k}}\right)\label{eq:proof_dyn_log_gain_sens}
\end{align}
\end{subequations}

\hrulefill{}
\end{figure*}
\bibliographystyle{IEEEtran}
\bibliography{arXiv/literature}

\end{document}